\pdfoutput=1
\documentclass{nature}

    \usepackage{authblk}
    \usepackage{amsmath,amssymb,amsfonts}
    \usepackage{mathptmx} % Times font for equations
    
    \usepackage{siunitx}
    \usepackage{svg}
    \usepackage{lipsum}

    \usepackage{graphicx} 
    \usepackage{xcolor}
    \usepackage[font=small,labelfont=bf]{caption}
    \usepackage{float}
    \usepackage[version=4]{mhchem}
    \usepackage[section]{placeins}

    \usepackage[resetlabels]{multibib}
    \newcites{sicite}{\_}%  \citelatex, \nocitelatex, ...

\newif\ifsubmit
\submittrue

\newcommand{\etc}{\textit{etc}}
\newcommand{\eg }{\textit{e.g. }}
\newcommand{\ie }{\textit{i.e. }}
\newcommand{\sub}[1]{\textsubscript{#1}}

\ifsubmit
    \definecolor{comments}{rgb}{0.1, 0.66, 0.1}
    \newcommand{\mao}[1]{}
    \newcommand{\can}[1]{}
    \newcommand{\yh}[1]{}
    \newcommand{\jps}[1]{}
    \newcommand{\todo}[1]{}
    \newcommand{\tocite}[1]{}
    \newcommand{\rev}[1]{#1}
    
\else
    \definecolor{comments}{rgb}{0.1, 0.66, 0.1}
    \newcommand{\can}[1]{[{\color{comments}CL: #1}]}
    \newcommand{\mao}[1]{[{\color{comments}M: #1}]}
    \newcommand{\yh}[1]{[{\color{comments}YH: #1}]}
    \newcommand{\jps}[1]{[{\color{comments}JPS: #1}]}
    \newcommand{\todo}[1]{[{\color{red}TODO: #1}]}
    \newcommand{\tocite}[1]{[{\color{red}citation-#1}]}
    \newcommand{\rev}[1]{{\color{blue}#1}}
    
\fi

    \PassOptionsToPackage{hyphens}{url}\usepackage{hyperref}

    \title{Experimentally realized memristive memory augmented neural network}
    \author[1]{Ruibin Mao}
    \author[1]{\rev{Bo Wen}}
    \author[1]{Yahui Zhao}
    \author[2,3]{Arman Kazemi}
    \author[3]{Ann Franchesca Laguna}
    \author[3]{Michael Neimier}
    \author[3]{X. Sharon Hu}
    \author[2]{Xia Sheng}
    \author[2,*]{Catherine E. Graves}
    \author[4,5,*]{John Paul Strachan}
    \author[1,*]{Can Li}
    
    \affil[1]{Department of Electrical and Electronic Engineering, The University of Hong Kong, Hong Kong SAR, China}
    \affil[2]{Hewlett Packard Labs, Hewlett Packard Enterprise, Milpitas, CA, USA}
    \affil[3]{Department of Computer Science and Engineering, University of Notre Dame, Notre Dame, IN, USA}
    \affil[4]{Peter Grünberg Institut (PGI-14), Forschungszentrum Jülich GmbH, Jülich, Germany}
    \affil[5]{RWTH Aachen University, Aachen, Germany}
    \affil[*]{canl@hku.hk, j.strachan@fz-juelich.de, catherine.graves@hpe.com}
    
    \date{}

    \makeatletter
    \let\newtitle\@title
    \let\newauthor\@author
    \makeatother
    
\begin{document}
    \maketitle

    \begin{abstract}
    \begin{center} Abstract \end{center}
    Lifelong on-device learning is a key challenge for machine intelligence, and this requires learning from few, often single, samples. 
    Memory augmented neural network has been proposed to achieve the goal, but the memory module has to be stored in an off-chip memory due to its size. 
    Therefore the practical use has been heavily limited. 
    Previous works on emerging memory-based implementation have difficulties in scaling up because different modules with various structures are difficult to integrate on the same chip and the small sense margin of the content addressable memory for the memory module heavily limited the degree of mismatch calculation. 
    In this work, we implement the entire memory augmented neural network architecture in a fully integrated memristive crossbar platform and achieve an accuracy that closely matches standard software on digital hardware for the Omniglot dataset. 
    The successful demonstration is supported by implementing new functions in crossbars in addition to widely reported matrix multiplications. 
    For example, the locality-sensitive hashing operation is implemented in crossbar arrays by exploiting the intrinsic stochasticity of memristor devices.
    Besides, the content-addressable memory module is realized in crossbars, which also supports the degree of mismatches. 
    Simulations based on experimentally validated models show such an implementation can be efficiently scaled up for one-shot learning on the Mini-ImageNet dataset. 
    The successful demonstration paves the way for practical on-device lifelong learning and opens possibilities for novel attention-based algorithms not possible in conventional hardware.
    \end{abstract}
    \pagebreak

    \section*{Introduction}
    
    Deep neural networks (DNN) have achieved massive success in data-intensive applications but fail to tackle tasks with a limited number of examples.  
    On the other hand, our biological brain can learn patterns from rare classes at a rapid pace, which could relate to the fact that we can recall information from an associative, or content-based addressing, memory. 
    Inspired by our brain, recent machine learning models, such as memory augmented neural networks (MANN)\cite{pmlr-v48-santoro16}, have adopted a similar concept, where explicit external memories are applied to store and retrieve the learned knowledge. 
    While those models have shown the ability to generalize from rare cases, they are struggled to scale up\cite{stevens2019manna, rae2016scaling}. 
    This is because, the entire external memory module needs to be accessed from the memory to recall the learned knowledge, which immensely increases the memory overhead.
    %to recall the learned knowledge from memory, the algorithm needs to access all the cells in the memory, which immensely increases memory overhead. 
    The performance is thus bottlenecked by the memory bandwidth and capacity issues\cite{strubell-etal-2019-energy, li2016evaluating, Ranjan2019XMANN} in hardware with the traditional von-Neumann computing architecuture\cite{von1993first}, especially when they are deployed in edge devices, where the energy sources are limited. 

Emerging non-volatile memories, e.g., memristors\cite{chua1971memristor}, have been proposed and demonstrated to solve the bandwidth and memory capacity issues in various computing workloads, including deep neural networks\cite{xia2019memristive, li2019long, wang2019situ, strukov2015nature, IBM2018equivalent, NTHU2019NE}, signal processing\cite{li2018analogue, Sheridan2017sparse}, scientific computing\cite{le2018mixed, zidan2018general}, solving optimization problems\cite{Yang2020optimization, Cai2020NE}, and more. 
Those solutions are based on the memristor's ability to directly process analog signals at the location where the information is stored. 
Most existing demonstrations mentioned above, however, \rev{mainly focus on executing matrix multiplications for accelerating deep neural networks with crossbar structures\cite{hu2016dot,chua1971memristor, xia2019memristive, li2019long, wang2019situ, le2018mixed, zidan2018general, strukov2015nature}, whose experience cannot be directly applied to the models with explicit external memories in MANNs.} 
Recently, several pioneering works aim to solve the problem with memristor-based hardware. 
One promising solution is to exploit the hyperdimensional computing paradigm \cite{Karunaratne2020NE}. 
But the high dimension vectors, as required to maintain the quasi-orthogonal nature, lead to excessive memory requirements. 
The recent prototype of this framework uses a large number of phase-change memory devices to solve the important yet simple Omniglot problem, the most commonly used handwritten dataset for few-shot image classification.\cite{karunaratne2021robust}. 
% \jps{The comparison to the IBM paper may be an important one - can we give any quantitative comparisons of their approach vs ours?  Could we make a table to contrast the approaches?  They also modified the algorithm to be HD compatible - do they get better error than us?}\can{they tried a 100-way problem as I remember. We can plan for en evaluation for the same problem but I would expect a similar performance in terms of accuracy...}
Although a close to software-equivalent accuracy was achieved, \rev{the number of devices required by the dimension of the binary feature vectors} will increase significantly for more complex problems. 
Ferroelectric device based ternary content addressable memory (TCAM) has been proposed to be used as the hardware to calculate the similarity directly in the memory\cite{Ni2019NE, Laguna2019GLSVLSI}, but it is only suitable for degree of mismatch up to a few bits. 
Besides, the locality sensitive hashing (LSH) function that enables the degree of mismatch search in the TCAM was implemented in software, and the experimental demonstration was limited to a 2$\times$2 TCAM array.  
More recently, a 2T-2R TCAM associative memory was used to demonstrate few-shot learning by calculating $L_1$ distance \cite{li2021sapiens}. In this work a 2-bit readout scheme is employed (requires 64 cycles per row) which incurs high energy and latency overheads, and feature extraction is again relegated to a digital processor.
The experimental demonstration of the entire MANN concept still remains a high risk, since the imperfections in such analog hardware, such as device variation, fluctuation, state drift, and readout noise during the parallel matrix multiplication operations, have challenged the feasibility of the promising hardware. 
    
    In this work, we experimentally demonstrate one- and few-shot learning with the {\em entire MANN} fully implemented in our integrated memristor hardware.
    To achieve this goal, we implement different functionalities in crossbars that are different from the widely reported matrix multiplication operations. 
    One enabler is the locality-sensitive hashing (LSH) function in crossbars by fully exploiting the intrinsic stochasticity of memristor devices.
    This is different from crossbars for matrix multiplications, where the stochasticity needs to be minimized. 
    Another innovation is implementing the search function and using the crossbar as a ternary content addressable memory (TCAM). 
    In addition to what's possible with conventional TCAMs, the proposed scheme can also measure the degree of mismatch reliably, which is crucial for few-shot learning implementation. 
    Since the requirements for those functions are different from conventional matrix multiplications, here we introduce several hardware-software co-optimization methods, including the introduction of the wildcard `X' bit in the crossbar-based LSH, and the careful choice of conductance range according to the device statistics.

    Finally, with a fully integrated memristor computing system, we are able to experimentally demonstrate the few-shot learning with a complete MANN model, including a five-layer convolutional neural network (CNN), the hashing function, and the similarity searches. 
    Taking into consideration all imperfections in the emerging system, our hardware achieves 94.9\% $\pm$ 3.7\%  accuracy in the 5-way 1-shot task with the Omniglot dataset\cite{lake2011one}, a popular benchmark for few-shot image classification, and 74.9\% $\pm$ 2.4\% accuracy in 25-way 1-shot task, which is close to the software baseline (95.2\% $\pm$ 2.6\% for 5-way 1-shot and 76.0\% $\pm$ 2.7\% for 25-way 1-shot). 
    Our experimentally-validated model also shows that the proposed method is scalable with a 58.7 \% accuracy to recognize rare cases (5-way 1-shot) for the Mini-ImageNet dataset\cite{vinyals2016matching}, where each image is a color (RGB) image of size 84$\times$84 pixels which is nine times larger than the size (28$\times$28) of images in the Omniglot dataset.
    This accuracy is only 1.3\% below the software baseline.  
    Owing to the in-memory analog processing capability and the massive parallelism, our hardware achieves more than 4,500$\times$ improvement in latency and \rev{2, 900$\times$ improvement in energy consumption on crossbar arrays} as compared to the general-purpose graphic processing unit (GPGPU) (Nvidia Tesla P100). 
    % 4.14 TOPS and consumes \SI{1.32}{\micro\joule} per image for Omniglot dataset from end to end. 

    \section*{Memory augmented neural networks in crossbars}

    % Fig. 1
    The MANN architecture commonly includes a controller, an explicit memory, and a content-based attention mechanism between the two. 
    A controller is usually a traditional neural network structure such as a convolutional neural network, a recurrent neural network (RNN), or the combination of different neural networks. 
    The explicit memory stores the extracted feature vectors as the key-value pairs so that the model can identify the values based on the similarity or distances between the keys. 
    The access of the explicit memory is the performance bottleneck for models that run on conventional hardware, such as the general-purpose graphic processing unit (GPGPU).
    It is because the similarity search requires accessing all the content in the memory; thus the repeated data transfer process delays the readout process and consumes abundant energy, especially when the memory needs to be placed in a separate chip. 
    
    We implement the entire network in our memristor-based hardware system to address the expensive data transfer issue in conventional digital hardware. The system aims to perform both matrix multiplication, for neural network controllers, and the similarity search, for the explicit memory, directly in the memristor crossbars, where the data are stored.
    Memristors have demonstrated great success in accelerating matrix multiplications in neural networks, which is also the controller part of the MANN, owing to their extreme density\cite{pi2019memristor}, the compatibility with CMOS integration\cite{cai2019fully,xia2019memristive}, and the capability to process analog information directly in its memory. 
    But, the demonstrations for explicit memories are still limited to small-scale experiments\cite{Ni2019NE, Laguna2019GLSVLSI} or \rev{simulated multiplication-and-add with readout conductances\cite{karunaratne2021robust}.}
    
\begin{figure}[htbp]
        \centering 
        \includegraphics[width=0.95\textwidth]{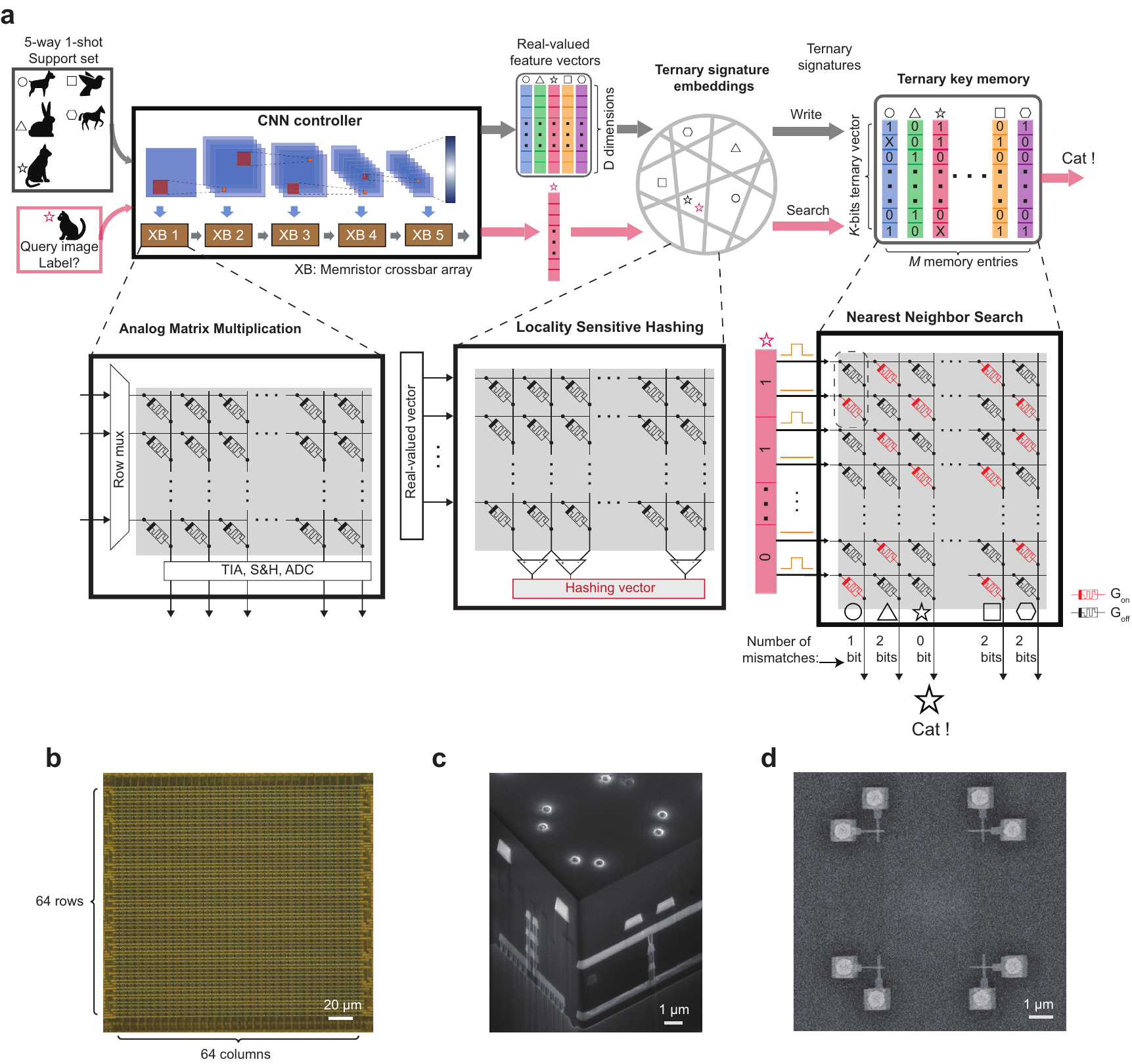}
        \caption{\textbf{Memory augmented neural networks in crossbar arrays.} 
        \textbf{a}, The schematic of a crossbar-based MANN architecture. 
        The expensive data transfer in a von Neumann architecture can be alleviated by performing analog matrix multiplication, locality sensitive hashing, and nearest neihgbor searching directly in memristor crossbars, where the data is stored. 
        \textbf{b}, Optical image of a 64$\times$64 crossbar array in a fully integrated memristor chip. 
        \textbf{c}, Cross-section of the memristor chip, where complementary metal-oxide-silicon (CMOS) circuits at the bottom, inter-connection in the middle, and metal vias on the surface for memristor integration with back-end processes.
        \textbf{d}, Top view of four 50 nm$\times$50 nm integrated cross-point memristors.
        Animal figures in \textbf{a} are taken from \url{www.flaticon.com}}
        \label{fig:MANN}
    \end{figure}

    Fig. \ref{fig:MANN}\textbf{a} illustrates how we implement the MANN in the crossbars. 
    First, a regular crossbar-based convolutional neural network is used to extract the real-valued feature vector, and the method implementing this step has been widely reported previously\cite{wang2019situ, yao2020fully, shafiee2016isaac}. 
    After that, distances are calculated between the extracted feature vector and those stored in a memory.
    Cosine distance (CD) is one of the most widely used distance metrics in the explicit memory of various MANN implementations, but it is not straightforward to implement with memristor-based crossbars. 
    On the other hand, the cosine distance between two real-valued vectors can be well approximated by the Hamming distance (HD) using locality sensitive hashing (LSH) codes of the two vectors\cite{kaiser2017learning, Ni2019NE, datar2004locality}.
    Accordingly, in this work, instead of being stored in a dynamic random access memory (DRAM) for future distance calculations, the features are hashed into binary/ternary signatures in a crossbar with randomly distributed conductance at each crosspoint exploiting the stochasticity of memristor devices.
    Those signatures are then searched against those previously stored in another crossbar that acts as a content-addressable memory enabled by a newly proposed coding method, from which we can also calculate the degree of mismatch that approximates the cosine distance of the original real-valued vector. 

    The idea is experimentally demonstrated in our integrated memristor chip. 
    One of the tiled \rev{$64\times64$} memristor crossbars in our integrated chip that we used to experimentally implement the network is shown in Fig. \ref{fig:MANN}\textbf{b}.
    The peripheral control circuits, including the driving, sensing, analog-to-digital conversions, and the access transistors, are implemented with a commercial 180 nm technology integrated chip (Fig. \ref{fig:MANN}\textbf{c}). 
    50 nm$\times$50 nm Ta/\ce{TaO_x}/Pt memristors are integrated with back-end-of-the-line (BEOL) processing on top of the control peripheral circuits (\ref{fig:MANN}\textbf{d}).
    The fabrication details, the device characteristics \rev{and peripheral circuit designs} were reported previously elsewhere\cite{sheng2019low, Li2020imw}, \rev{and the picture of our test chip and platform is shown in Supplementary Fig. \ref{fig:setup}}.

    \section*{Locality sensitive hashing in crossbar array}
    
\begin{figure}
        \begin{center}
        \includegraphics[width=0.95\textwidth]{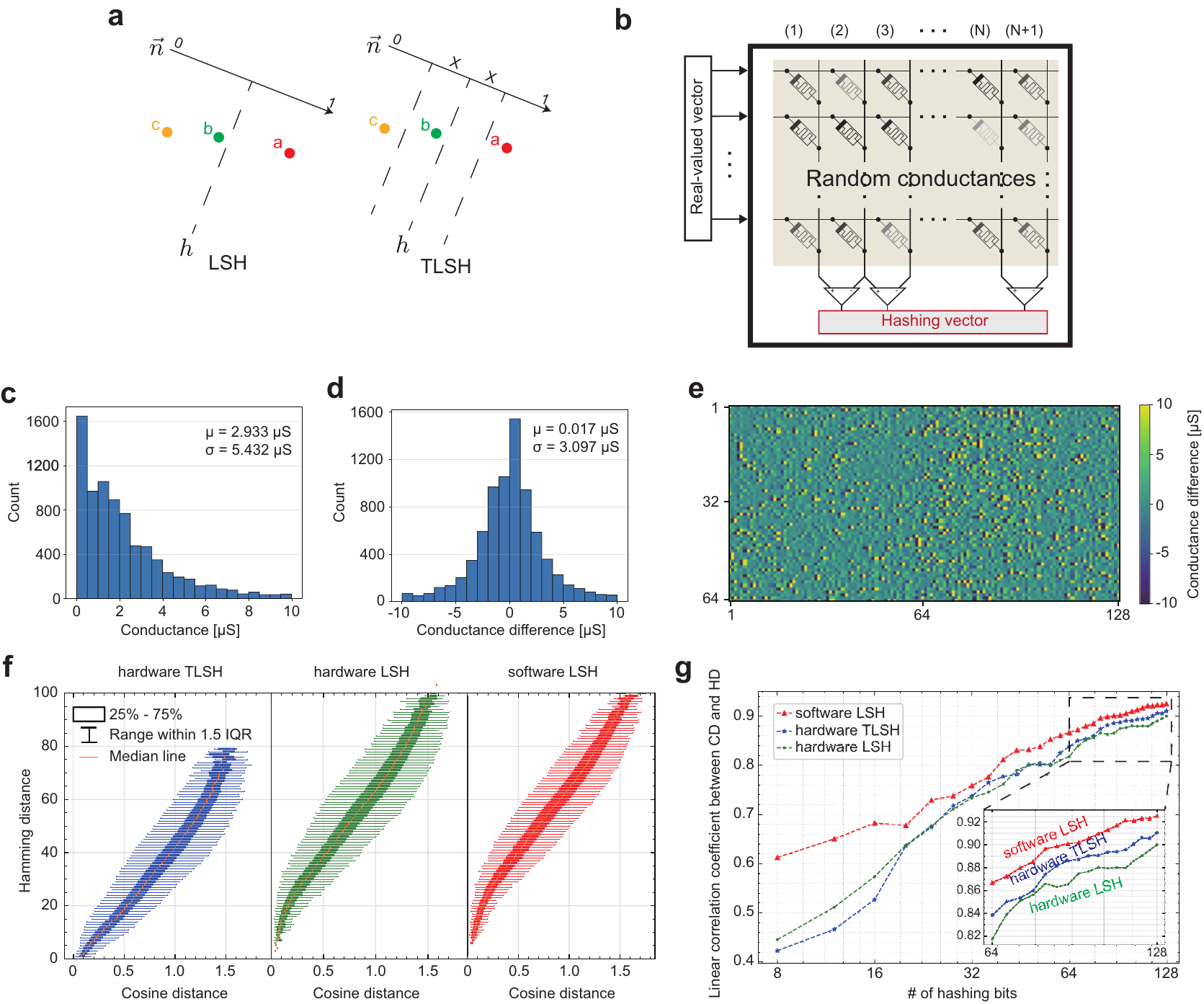}
        \caption{\textbf{Robust ternary locality sensitive hashing in analog memristive crossbars.} 
        \textbf{a}, Illustration of the Locality Sensitive Hashing (LSH) and the Ternary Locality Sensitive Hashing (TLSH) concept. 
        \textbf{b}, The LSH or TLSH implemented in memristor crossbars. 
        Each adjacent column pair represents one hashing plane. 
        So, crossbars with $N+1$ columns can generate $N$ hashing bits with this method.
        Greyscale colors on the memristor symbol represent random conductance states.
        \textbf{c}, A random memristor conductance distribution in a 64$\times$129 crossbar after applying five RESET pulses to each device. The intrinsic stochastic behavior in memristor devices results in a lognormal-like distribution near \SI{0}{\micro\siemens}.
        \textbf{d}, The distribution of the memristor conductance difference for devices in adjacent columns.  
        The differential conductance distribution is random with zero-mean, matching the requirements of our hashing scheme. 
        \textbf{e}, The conductance difference map of size 64$\times$128 (including three crossbar arrays each of size 64$\times$64).
        \textbf{f}, The correlation between cosine distance and Hamming distance with different hashing representations shows that the Hamming distance generated by both hardware and software can well approximate the cosine distance. IQR, interquartile range. 
        \textbf{g}, The linear correlation coefficient between Hamming distance and cosine distance increases with the number of total hashing bits. 
        The hardware TLSH approach shows a higher correlation coefficient than the hardware LSH approach due to the reduced number of unstable bits, as detailed in Supplementary Fig. \ref{fig:bitflip}. 
        }
        \label{fig:TLSH}
        \end{center}
    \end{figure}

    We first introduce the idea of performing locality sensitive hashing (LSH) directly in crossbars employing intrinsic stochasticity, and validate that the experimental implementation in memristor crossbars can approximate cosine distance.     
    LSH \cite{gionis1999similarity, shinde2010similarity, huang2015query} is a hashing scheme that generates the same hashing bits with a high probability for the inputs that are close to each other. 
    One of the hashing functions among the LSH family is implemented by random signed projections, \ie, applying a set of hyperplanes to divide the Hilbert space into multiple parts such that similar inputs project to the same partition with a high probability (Fig. \ref{fig:TLSH}\textbf{a}). 
    This random projection is mathematically expressed by a dot product of the input vector $\vec{a}$ and a random normal vector $\vec{n}$, so that `1' is generated if $\vec{a}\cdot\vec{n}>0$, and `0' otherwise. 
    Accordingly, LSH bits can be calculated by Equation \ref{eq:lsh} in a matrix form,
    \begin{equation}
    \label{eq:lsh}
    \vec{h}=H\left(\vec{a}~\mathbf{N}\right)
    \end{equation}
    where the $\vec{h}$ is the binarized hashing vector, $\vec{a}$ the input real-valued feature vector, $\mathbf{N}$ the random projection matrix with each column a random vector, and $H$ the Heaviside step function.

    The random projection matrix can be constructed physically by exploiting the stochastic programming dynamics of the memristor devices or the initial randomness after the fabrication (Fig. \ref{fig:TLSH}\textbf{b}).
    But it is still challenging to generate random vectors $\vec{n}$ with a zero mean value, as required by the LSH algorithm, because the conductance of the memristor device can only be positive values. 
    Our solution is to take the difference between devices in the adjacent columns\cite{kazemi2020device} in the crossbar array. 
    The devices from the columns, assuming no interference in between, are independent of each other.
    Therefore, the distribution of the conductance difference will also be uncorrelated and random. 
    In this way, the random normal vector $\vec{n}$ in the original equation can be represented by the difference of two conductance vectors, \ie $\vec{n} = \left(\vec{g^+}-\vec{g^-}\right)/k$. 
    The zero mean value of the vector $\vec{n}$ is guaranteed as long as the distribution of the memristor conductance vectors ($\vec{g^+},\vec{g^-}$) have the same mean value. 
    $k$ is a scaling factor, which can be set as an arbitrary value because we only need to determine if the output is larger than zero or otherwise.
    
    % Experiment
    Here, we experimentally program all devices in a crossbar array to the same target conductance state.
    The programming process of memristor devices is stochastic\cite{xia2019memristive} as the thinnest part of the conductive filament can be only a few atoms wide\cite{niu2016geometric}. 
    Accordingly, the final conductance values follow a random distribution with the mean roughly matching the target conductance. 
    To lower the output current and thus the energy consumption, we reset all devices \rev{close to the highest resistance state that we can achieve (\SI{17}{\nano\siemens} at the read voltage of \SI{0.2}{\volt}) from arbitrary initial states using a few pulses (see Methods for details).} 
    After programming, as expected, \rev{most devices are programmed to a conductance state near the highest resistance state} (Fig. \ref{fig:TLSH}c), and the difference between devices from adjacent columns follows a random distribution with a zero mean value (Fig. \ref{fig:TLSH}\textbf{d, e}).
    Hashing bits for an input feature vector are generated efficiently by performing multiplications with the randomly configured memristor crossbar array. 
    After converting the real-valued input vector into the analog voltage vector, the dot product operations are conducted by applying the voltage vectors to the row wires of the crossbar and reading the current from the column wires.
    Thus, the hashing operation is completed by comparing the current amplitude from the adjacent columns (Fig. \ref{fig:TLSH}\textbf{b}) in one step, regardless of the vector dimension. 
    
\begin{figure}[!hth]
        \centering 
        \includegraphics[width=0.95\textwidth]{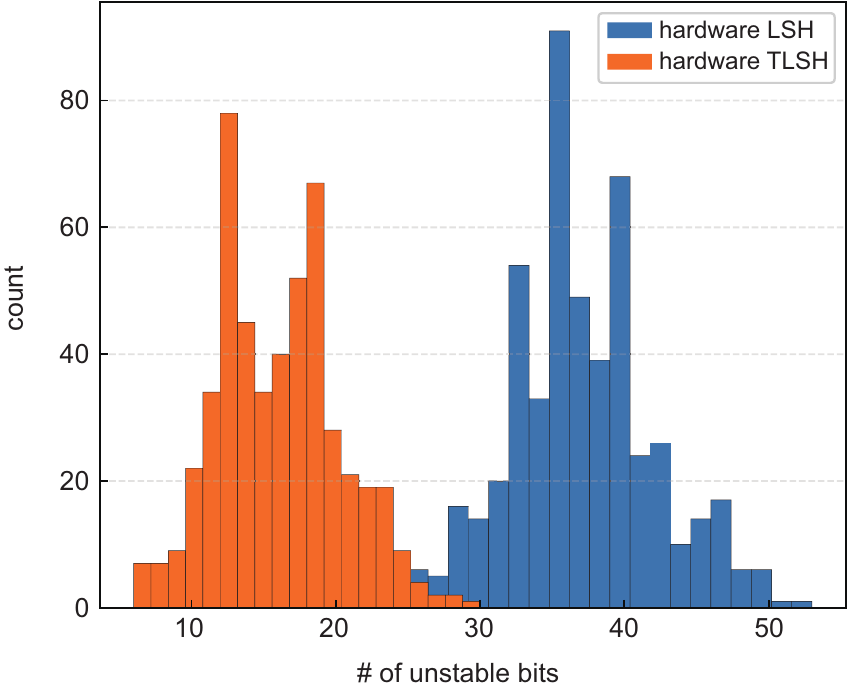}
        \caption{\textbf{Number of unstable bits using hardware TLSH and LSH.} 
        Ideally, we expect no unstable bits when performing hardware LSH so that it's equal to software LSH. However, due to many device nonidealities, there will be many unstable bits mainly caused by the conductance fluctuation. In this work, we apply TLSH to solve this problem. Here, we use the same set of vectors in the main text and repeat the hashing process 100 times with the same conductance map of size $64\times129$. We then count the number of unstable bits 100 times for both TLSH and LSH. The result shows that by applying TLSH we can reduce the number of unstable bits, thus increasing the robustness of our hardware in terms of hashing process.}
        \label{fig:bitflip}
    \end{figure}

    Imperfections in emerging memristive devices, such as conductance relaxation and fluctuation, limit experimental performance. 
    This is mainly because the device conductance fluctuations incur instability of hashing planes implemented as adjacent column pairs in crossbar arrays. 
    This causes hashing bits for input vectors that are close to hyperplanes to flip between 0 and 1 over time (Supplementary Fig. \ref{fig:bitflip}) and therefore leads to an inaccurate approximation of the cosine distance. 
    % To our knowledge, there is no report of experimental implementations of LSH in crossbar arrays with device imperfections and noise from peripheral circuits taken into consideration. 
    It should be emphasized that this problem only becomes apparent when performing matrix multiplication in arrays, rather than multiplying with the readout conductances reported in other works\cite{karunaratne2021robust}. 

    To mitigate bit-flipping, we propose a software-hardware co-optimized ternary locality sensitive hashing scheme (TLSH). 
    The scheme introduces a wildcard `X' value to the hashing bits (Fig. \ref{fig:TLSH}\textbf{a}), in addition to `0' and `1' in the original hashing scheme. 
    As the name implies, the Hamming distance between the wildcard `X' and any values will always be zero.
    The `X' is generated when the absolute value of the output current difference is smaller than a threshold value, \ie $I_\text{th}$.
    The value for $I_\text{th}$ is chosen to be small and close to the transient analog computing error from the crossbar, such that any bit-flipping minimally impacts the similarity search.

    We validate the proposed approach by conducting experiments in our integrated memristor crossbars.
    The hash outputs for 500 64-dimensional real-valued random vectors are computed in our memristor crossbars for binary and ternary hashing vectors. 
    The $I_\text{th}$ representing the threshold of the `X' wildcard is set to \SI{4}{\micro\ampere} in the ternary hashing implementation (TLSH). 
    Fig. \ref{fig:TLSH}\textbf{f} shows that the cosine distance is closely correlated with the Hamming distance between the hashed vectors with 128 hashing bits in total, regardless of whether the hashing codes are generated by a 32-bit floating-point digital processor (``software LSH'' in Fig. \ref{fig:TLSH}\textbf{f}), the analog crossbar (``hardware LSH''), or the proposed co-designed ternary hashing codes by the crossbar (``hardware TLSH''). 
    Note that the Hamming distance of the ternary hashing codes is smaller than that of the binary codes because the distance to a wild card `X' is always zero. 
    The effectiveness of the method is evaluated quantitatively by the linear correlation coefficient, as shown in Fig. \ref{fig:TLSH}\textbf{g}.
    The result shows that the proposed ternary hashing narrows the already small gap between the digital software approach and our analog hardware approach.  
    The performance improvement results from significantly reduced unstable bits, which is experimentally demonstrated by the comparison shown in Supplementary Fig. \ref{fig:bitflip}. 
    The results demonstrate that crossbar arrays, utilizing the proposed ternary scheme, can effectively and efficiently perform hashing operations, taking advantage of intrinsic stochasticity and massively parallel in-memory computing. 

    % \section*{Ternary Hamming distance computed as dot product in crossbar arrays}
    \section*{TCAM in crossbars with ability to output degree of mismatches}
    
    Following the LSH step, the binary or ternary hashing signatures will be searched against the hashed signatures previously stored in a memory to calculate the similarity and thus find the $k$-closest match. 
    As mentioned earlier, this is an extremely time- and energy-consuming step on conventional hardware such as GPUs.
    Content addressable memories (CAM) or the ternary version (TCAM) are direct hardware approaches that can find the exact match in the memory in one step. 
    Still, existing static random access memory (SRAM) based CAM/TCAM implementations limit the available memory capacity and incur high power consumption.
    CAMs/TCAMs based on non-volatile memories have been developed recently, including those based on memristor/ReRAM (\eg 2T-2R\cite{li20131, li2021sapiens}, 2.5T-1R\cite{lin20167}), floating gate transistor (\eg 2Flash\cite{fedorov2014area}), ferroelectric transistors (\eg 2FeFET\cite{Ni2019NE, Laguna2019GLSVLSI}), \etc.  
    Although these studies demonstrated good energy efficiency, they are limited to at most a few bits mismatches which have difficulties to serve as attentional memory modules for scaled-up MANNs\cite{Ni2019NE, Laguna2019GLSVLSI, li2021sapiens}. 
    
\begin{figure}[htbp]
        \centering 
        \includegraphics[width=0.95\textwidth]{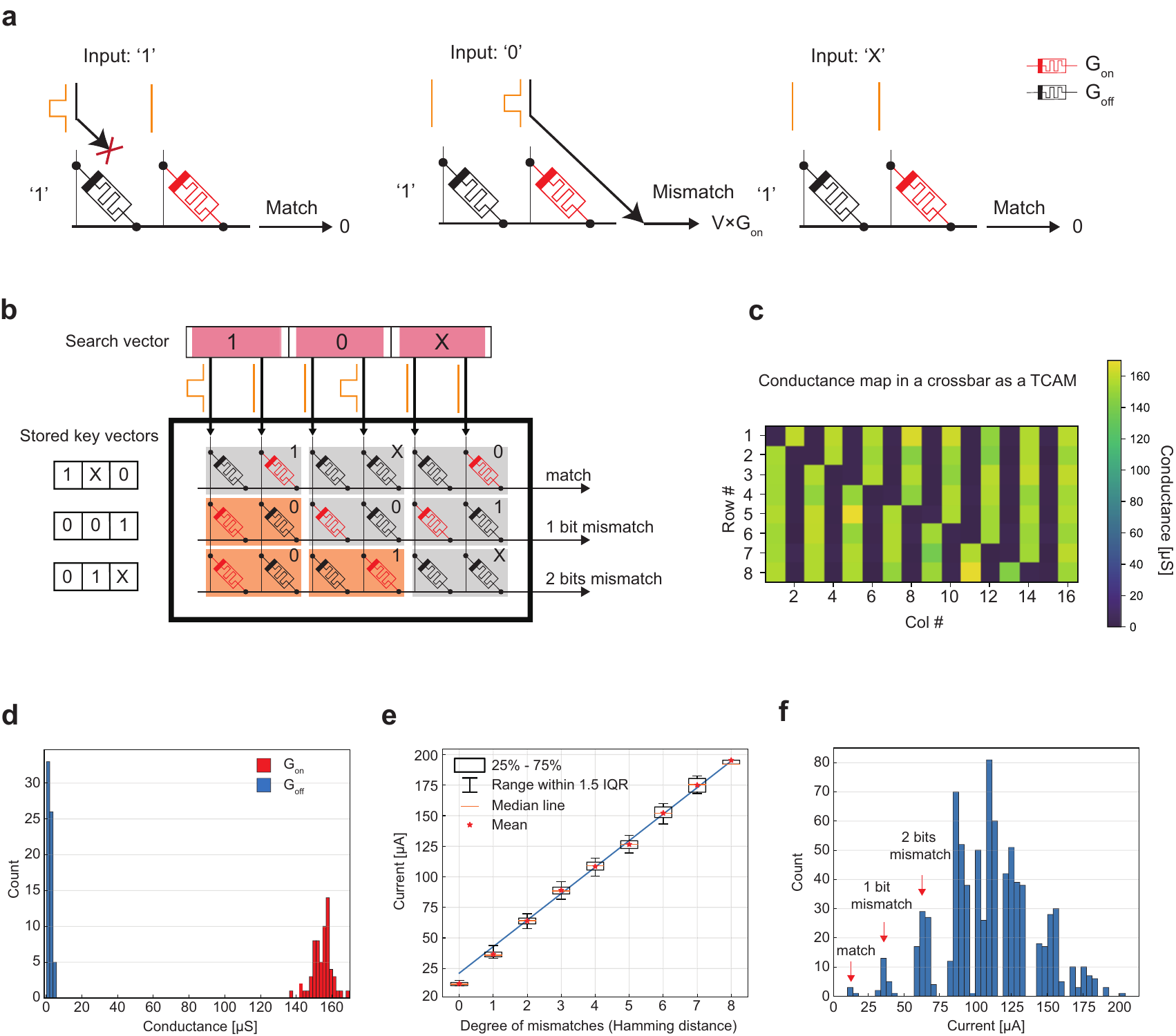}
        \caption{\textbf{TCAM implemented in crossbar array capable of conducting Hamming distance calculation.} 
        \textbf{a}, Illustration of the basic principle for using dot product to distinguish `match' and `mismatch' cases. 
        \textbf{b}, The schematic of calculating Hamming distance in a crossbar. 
        The figure shows three 3-dimensional ternary key vectors stored in a $3\times6$ crossbar with a differential encoding. 
        Differential voltages representing ternary bits in search vectors are applied to the source line and the output current from the bit line can represent the THD between the search vector and keys stored in the memory. 
        \textbf{c}, The readout conductance map after eight binary vectors experimentally stored in the crossbar as memory. 
        In the experiment, we set G\sub{on} as \SI{150}{\micro\siemens} and V\sub{search} as \SI{0.2}{\volt}. 
        \textbf{d}, Distribution of G\sub{on} and G\sub{off}. 
        \textbf{e}, Ouput current shows a linear relation with Hamming distance measuring the degree of mismatches. IQR, interquartile range. 
        \textbf{f}, Current distributions are separated from each other through which we can obtain the number of mismatch bits (\ie, Hamming distance). }
        \label{fig:TCAM}
    \end{figure}

    We implement the TCAM functionality directly in an analog crossbar with the additional ability to output the degree of mismatch based on the Hamming distance, rather than only a binary match/mismatch.
    In contrast to conventional TCAM implementations which sense a mismatch by a dis-charged match-line, our crossbar-based TCAM searches through a simple encoding and a set of dot product operations computed in the output currents. 
    Fig. \ref{fig:TCAM}\textbf{a} shows a schematic on how this scheme works. 
    First, the query signature is encoded to use a pair of voltage inputs for 1-ternary-bit, so that one column wire is driven to a high voltage (\ie V\sub{search}), while the other is grounded. 
    The corresponding memristor conductances that store previous signatures are encoded with one device set to a high conductance state (\ie, G\sub{on}) and the other to the lowest conductance state (\ie, G\sub{off} $\approx 0$).
    In this way, for a `match' case, the high voltage will be applied to the device in the low conductance state, and therefore, very little current is added to the row wires. 
    In a `mismatch' case, the high voltage applied to the device in the high conductance state will contribute V\sub{search}$\times$G\sub{on} to the output current of the column wires.
    The wildcard `X' in the ternary implementation will be naturally encoded as two low voltages as input or two low conductance devices so that they contribute zero or very little current and thus always yield `match'.
    In this way, the degree of mismatch, the Hamming distance, between the query signature and all words stored in the crossbar is computed in a constant time-step by sensing the column currents from the crossbar (see Fig. \ref{fig:TCAM}\textbf{b}).

    We have experimentally implemented the above TCAM for measuring Hamming distance in memristive crossbars. 
    First, eight different binary signatures, each has eight bits but different number of `1's (from one `1' to eight `1's), are encoded into conductance values as shown in Fig. \ref{fig:TCAM}\textbf{c}.
    % Eight binary vectors, 00000001, 00000011, 00000111, 00001111, 00011111, 00111111, 01111111, 11111111\can{I understand it is a little late, but we did not try the storing `X' case.}, are sequentially stored in crossbar array as shown in Fig. \ref{fig:TCAM}\textbf{a}. 
    The conductance values are then programmed to a crossbar with an iterative write-and-verify method (see ref.\cite{sheng2019low} and Methods for details), with the readout conductance matrix after successful programming shown in Fig. \ref{fig:TCAM}\textbf{b}. 
    We choose \SI{150}{\micro\siemens} as the G\sub{on} for a higher on/off conductance ratio and minimal relaxation drift (Supplementary Fig. \ref{fig:Grelax}). 
    Fig. \ref{fig:TCAM}\textbf{d} shows both the distribution of G\sub{on} and G\sub{off} after programming. 

    After configuring the memory to store the previously generated signatures, 100 ternary signature vectors as queries are randomly chosen and the corresponding encoded voltages are applied to the column wires of the crossbar (Fig. \ref{fig:TCAM}\textbf{c}), to perform the search operation. 
    The search voltage (V\sub{search}) is chosen to be \SI{0.2}{\volt} in this work, so each mismatched bit will contribute approximately \SI{30}{\micro\ampere} (=\SI{0.2}{\volt}$\times$\SI{150}{\micro\siemens}) to the output current. 
    %\can{explain why the experiment is not matching the math:}
    %\can{Here is }
    In the experiment, however, results are deteriorated by non-ideal factors. 
    For example, the memristor in a low conductance state still contributes a small current (V\sub{search}$\times$G\sub{off}$\neq$0) in a match case, imperfect programming of G\sub{on} results in deviations in output current for each `mismatch' case, \etc. 
    Our device exhibits a large enough on/off ratio\cite{sheng2019low}, but for devices with a lower conductance on/off ratio, such as MRAM\cite{apalkov2016magnetoresistive}
    % FeFET\cite{Laguna2019GLSVLSI,Reis2021ASPDAC,Ni2019NE,yurchuk2016charge}
    , the problem would be more significant. 
    For such cases, we propose a 3-bit encoding that is discussed in detail in Supplementary Fig. \ref{fig:3dpe}.
    Additionally, non-zero wire resistances cause voltage drop along wires, lowering the output current from what would be expected ideally. 
    Despite these factors, the output current in our experiments exhibits a linear dependence on the number of mismatch bits, \ie ternary Hamming distance (Fig. \ref{fig:TCAM}\textbf{e}). 
    Fig. \ref{fig:TCAM}\textbf{f} shows separated distributions where each distribution represents a distinct number of mismatch bits ranging from 0 to 8. 
    We have thus experimentally demonstrated a robust capability to store patterns, search patterns, and obtain the degree of mismatch which will enable determining the closest match that is stored in an array by simply comparing output currents. 

    \section*{One- and few-shot learning experiment fully implemented in memristor hardware}
    
    We implemented a complete MANN demonstrating one-shot and few-shot learning in crossbars. 
    To evaluate and compare the performance of our method, we chose the Omniglot dataset\cite{lake2011one}, a commonly used benchmark for few-shot learning tasks. 
    In this dataset, there are 1,623 different handwritten characters (classes) from various alphabets. 
    Each character contains 20 handwritten samples from different people. 
    Samples from 964 randomly chosen classes are used to train the CNN controller and the remaining 659 are used for one-shot and few-shot learning experiments. 
    In an $N$-way $K$-shot learning experiment, the model should be able to learn to classify new samples from $N$ different characters (classes) after being shown $K$ handwritten images from each character (support set). 
    The accuracy is evaluated by classifying an unseen sample (query set) after learning from the limited number of samples (only one sample each for the 1-shot problem) from each class.

    \begin{figure}[htbp]
        \centering 
        \includegraphics[width=0.95\textwidth]{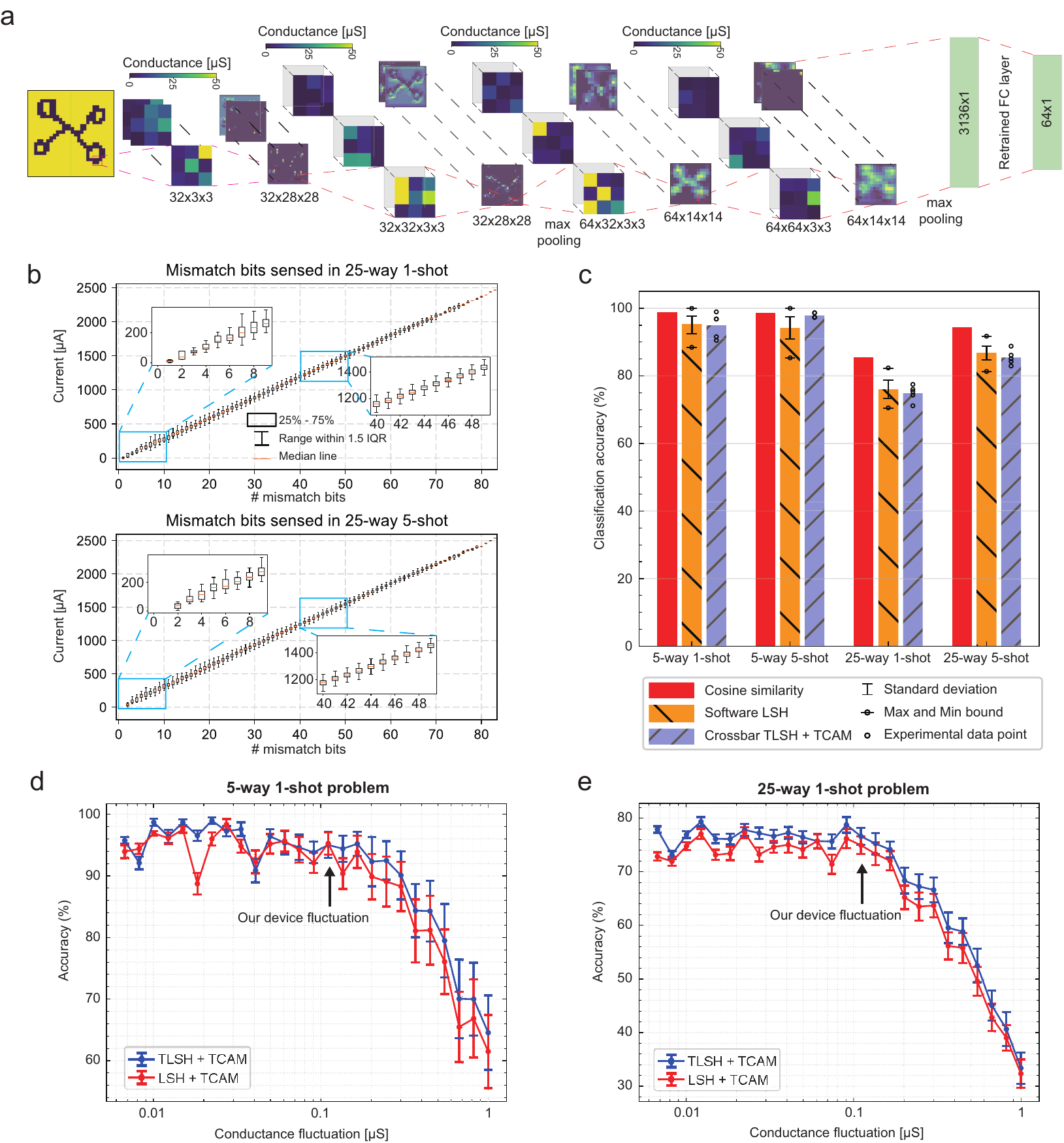}
        \caption{
            \linespread{1.0} \selectfont{}
        \textbf{End-to-end experimental inference with memristive crossbar arrays.}
        \textbf{a}, Schematic of CNN structure implemented in the memristor crossbar array. The conductance shows the weight mapping of CNN kernels. The format of dimension representations in the figure follows the Output channel (O), Input channel (I), Height (H), and Width (W). The conductance maps representing the whole CNN kernels are shown in Supplementary Fig. \ref{fig:cnngmap}. 
        \textbf{b}, \rev{Linear relationship between the sensing current from the crossbar-based TCAM and the number of mismatch bits during the search operations.} 
        \textbf{c}, Classification accuracy with cosine similarity, software-based LSH with 128 bits, and end-to-end experimental results on crossbar arrays. We provide 5 experimental data points for each task. Software LSH shows experimental variation due to different initializations of the hashing planes in each experiment.
        %\textbf{d}, Comparison over LSH and TLSH with both simulation and experiment on few shot learning tasks.
        %\textbf{e} Average energy consumption of each TCAM search operation with TLSH and LSH. 
        \textbf{d, e}, Simulations of classification accuracy of 5-way 1-shot problem (\textbf{d}) and 25-way 1-shot problem (\textbf{e}) as a function of device fluctuations in the memristor model for both TLSH and LSH. Fluctuations from nearly zero to \SI{1}{\micro\siemens} are shown. The actual experimental fluctuation level is shown with an arrow.
        }
        \label{fig:CNN}
    \end{figure}

    In our experiment, the memristor CNN controller first extracts the feature vector from an image. Note that the CNN weights don't need to be updated after the meta-training process which is done in software offline. 
    The CNN consists of four convolutional layers, two max-pooling layers, and one fully connected layer (Fig. \ref{fig:CNN}\textbf{a}).
    There are nearly 65,000 weights in convolutional layers altogether that are represented by \rev{130,000} memristors, with the conductance difference of two memristors representing one weight value. 
    The weights of convolutional layers are flattened and concatenated first (see Supplementary Fig. \ref{fig:cnnlayer} and Methods for details) and then programmed to crossbar arrays with an iterative write-and-verify method. 
    Limited by the available array size, we divide larger matrices into 64$\times$64 tiles and reprogram the same arrays when needed to accomplish all convolutional operations in the crossbar. 
    Experimental conductance maps (36 matrices of conductance values) for the CNN layers after each programming of an array are shown in Supplementary Fig. \ref{fig:cnngmap}.
    The repeated programming of memristor arrays demonstrated good reliability of the memristor devices within crossbars.
    After programming the convolutional weights into the crossbar, the fully connected layer is retrained to adapt to the hardware defects. 
    
    Before the few-shot learning, the explicit memory stored in the crossbar-based TCAM is initialized with all `0's. 
    During the few-shot learning, the feature vectors computed by the memristor CNN next hashed into 128 binary or ternary signatures and searched against the entries in the crossbar-based TCAM for the closest match, with the methods described before. 
    The label of the closest match will be the classification result.
    If correct, the nearest neighbor entry will be updated based on the new input query vector (see Methods). 
    Otherwise, the signature along with the label is written to a new location in the TCAM using differential encoding. 
    \rev{
    % \can{I don't quite understand the difference here - 
    % Different from the memory update algorithm for life-long learning proposed in the previous work\cite{kaiser2017learning}, our proposed binary update algorithm directly update the binary vector stored in the crossbar-based TCAM. This allows efficient in-place updating of the memory by updating the conductance of memristors through a few voltage pulses\cite{li2018efficient}.
    After learning $K$ images from the support set, The conductance map that is stored in the crossbar-based TCAM is shown in Supplementary Fig. \ref{fig:tcam_gmap}. 
    Note that the CNN controller stays the same for all four few-shot learning tasks, a key feature to support lifelong learning, which is different from the previously reported few-shot learnings based on high-dimensional computing\cite{karunaratne2021robust}. 
    The memory is the only part need to be updated during lifelong learning. 
    Even for the memory module, the update is not frequent (1.3 times per bit for 20-shot) throughout the learning process, as demonstrated in Supplementary Fig. \ref{fig:binary_updates}. 
    Considering a common statistical endurance (about $10^{6}$) for a relative high switching window reported previously\cite{zhao2018characterizing, zhao2019endurance}, we estimate about 4000-years lifetime of a face recognition system that sees 10 times each face in one day which is much longer than the human life.
    % , so that the endurance performance of memristors ($>10^{11}$ write cycles\cite{lee2011endurance, jiang2016sr}) well meets the requirement from lifelong learning. 
    }

    % \rev{
    % To further demonstrate that our crossbar-based TCAM can achieve wide sensing range and high sensing margin, we plot the output current of each search operation between every query and key in 25-way tasks in Fig. \ref{fig:CNN}\textbf{b}. 
    % The figure shows good linearity up to 80 mismatch bits and high sensing margin either in low hamming distance range or high hamming distance range. 
    % We also demonstrate that our TLSH method can provide a better sense margin compared with conventional LSH techniques without degrading the accuracy (see Supplementary Figure. \ref{fig:sense_margin}). Detailed discussion about sense margin can be found in Supplementary Note \ref{sinote:sense_margin}.
    % }

    Accuracy is evaluated experimentally in classifying new samples after few-shot learning with four standard tasks: 5-way 1-shot, 5-way 5-shot, 25-way 1-shot, and 25-way 5-shot, respectively. 
    \rev{
    We found that the experimental sensing currents during few-shot learning experiment are highly linear with the number of mismatch bits, \ie hamming distance, as shown in Fig. \ref{fig:CNN}\textbf{b}.
    This is partly enabled by the introduced wildcard `X' from our TLSH method, as discussed in detail in Supplementary Fig. \ref{fig:sense_margin} and Supplementary Note \ref{sinote:sense_margin}.}
    % The classification accuracy of the end-to-end experiment from physical crossbar arrays is shown in Fig. \ref{fig:CNN}\textbf{c}.
    % , along with the results from the digital hardware that implements the exact the same structure 
    %  the CNN controller and cosine distance search or LSH plus Hamming distance search using 128 hashing bits. 
    The classification results shown in Fig. \ref{fig:CNN}\textbf{c} demonstrate that for 5-way problems, our full crossbar hardware-based MANN achieves an accuracy very close to the software baseline implemented in digital hardware with cosine similarity as the distance metric.
    For 25-way problems, we find no difference between results from our analog hardware and that from the digital hardware implementing the same LSH plus Hamming distance algorithm. 
    Though there exists some accuracy drop compared to the cosine baseline, the performance can be improved to match the baseline accuracy by increasing the number of hashing bits from 128 to 512 (see Supplementary Fig. \ref{fig:multi_lsh}). 
    %there is a small degradation in the accuracy ($<$10\%) compared with cosine baseline \can{this is a lot, may compare with software LSH, or change another way to say it.}, but it is because of the limited number of hashing bits (128-bit). 
    %If we compare the results from the analog hardware with that from the digital hardware that implements the same LSH plus Hamming distance algorithm, no difference is found.
    %In addition, if we add the number of hyperplanes to 512, we can achieve the same accuracy as the cosine baseline (see Supplementary Fig. \ref{fig:multi_lsh}). 
    The experimental results demonstrate that the MANN fully implemented in crossbar arrays can achieve similar accuracy as software for this task. 
    %\can{we can't prove the energy and latency yet...should belong to the next section --- but have much lower energy consumption and time latency. }
    
    %\can{I'd prefer not to compare with the `RRAM model' here. One story at a time. --
    %In addition, results of simulation with our RRAM model match the experiment well on all few-shot learning tasks which proves our thoughts about how the nonidealities of memristors 
    %affect the classification accuracy.}

    %\can{I feel the following two paragraphs do not belong to the main text}
    %We then show in Fig. \ref{fig:CNN}\textbf{d} the classification accuracy of our hardware using both the TLSH and LSH as the hashing scheme. For each of these problems, TLSH exhibits better performance \can{higher classification accuracy? If so, be explicit} for both the simulation and hardware implementation especially when tackling more complex problems. 
    %More importantly, as mentioned before, introducing 'X' by applying TLSH in the experiment can further reduce the search energy as shown in Fig. \ref{fig:CNN}\textbf{e} since the current addition is zero for match cases where the input is 'X' which is illustrated in Fig. \ref{fig:TCAM}\textbf{b}. 
    %We finally achieved 13.8\% reduction in energy per search operation using TLSH. Note that the energy consumption can be further reduced for more complex problem and scaled-up MANNs\can{why?}, which is an essential problem for edge computing.
    \section*{Device imperfections analysis}
    % \jps{I think much or all of this paragraph could go to the methods or stay in the supplementary information}\can{I prefer to keep it here, because in my view it is essential for this experimental demonstration work.}
    The accuracy can be affected by many non-idealities in emerging memory devices, among which the two most prominent are conductance fluctuations and relaxation. 
    We noted that the conductance of memristor devices fluctuates up and down (see Supplementary Fig. \ref{fig:fluc}\textbf{a, b}) even within a very small period (at the scale of nanoseconds).
    The fluctuation leads to frequent changes in convolutional kernels, hyperplane locations in LSH operations, and stored signature values in TCAMs, which negatively impact the accuracy.
    % The change which is comparable to the time (100ns) between successive input vectors when conducting LSH in real hardware, causing a frequently changing of hyperplane locations. 
    The data (shown in Supplementary Fig. \ref{fig:fluc}\textbf{c}) measured from our integrated array shows that the degree of device fluctuation increases with the conductance value. 
    This behavior is consistent with previous reports on single device measurement\cite{sheng2019low, ambrogio2014statistical}. 
    In addition to the conductance fluctuations, the programmed value may also change permanently (relaxation) over time, which is characterized in detail in Supplementary Fig. \ref{fig:Grelax}\textbf{c}.
    From these results, we find that conductance relaxation is larger when device conductance is programmed to a certain range (from around \SI{25}{\micro\siemens} to \SI{75}{\micro\siemens}).
    Therefore, in our implementation, we try to avoid this range as much as possible to achieve the software equivalent accuracy.
    For example, in the LSH part, we chose lower conductance levels to minimize both conductance fluctuation and relaxation. 
    In the TCAM part, we chose \SI{0}{\micro\siemens} and \SI{150}{\micro\siemens} as the low and high conductance levels to minimize the impact of conductance relaxation. 
    In the CNN part, we observe that most weight values are very small (near zero), so with the differential encoding method (details described in Methods) we can guarantee that most memristor conductance values are below the range with higher relaxation.

    To analyze our software-competitive accuracy results and evaluate if our method is scalable, we built an empirical model describing experimental conductance-dependent fluctuation behavior and deviation after programming. 
    With the experimental calibrated model (see Methods and Supplementary Fig. \ref{fig:fluc}\textbf{c, d} for more details), we can match the simulation results with the experiments in few-shot learning on Omniglot handwritten images.
    The detailed comparison is shown in Supplementary Fig. \ref{fig:tlsh_lsh}\textbf{a}). 
    The simulation also enables us to analyze how different device fluctuations impact classification accuracy. 
    We conducted simulations assuming the device conductance fluctuation spanning from nearly no fluctuation to \SI{1}{\micro\siemens} which is about ten times larger than our device behavior. 
    The results in Fig. \ref{fig:CNN}\textbf{e, f} show that with the experimental fluctuation value, the accuracy stays almost the same as the software equivalent value, but the accuracy will sharply drop if the fluctuation is above three-times larger than our experimental value. 
    % For different fluctuation, we choose different threshold currents for TLSH (see Methods) based on experience and experimental data (see Supplementary Fig. \ref{fig:tlsh_x}). 
    % The results show that the crossbar-based MANN can only tolerate the conductance fluctuation up to as 3 times as our devices. Again, 
    The results also show that our proposed TLSH method exhibits better performance compared to the conventional LSH, especially for more significant device fluctuation scenarios (Fig. \ref{fig:CNN}\textbf{e, f} and Supplementary Fig. \ref{fig:acc_var}). 
    % We also explore the tolerance of device-to-device variation level which is shown in Supplementary Fig. \ref{fig:acc_var}. 
    In addition to the higher tolerance to the device fluctuation, the comparison shown in Supplementary Fig. \ref{fig:tlsh_lsh}\textbf{b, c} also demonstrates the TLSH's advantages in search energy. 
    % We can infer that more accurate approximation of cosine distance by TLSH indeed brings better results and introducing wildcard `X' will decrease the search energy. 
    These simulations, with experimental calibration, elucidate the experimentally observed defect tolerance and software-equivalent accuracy. 
    \rev{
    Though there exist other defects such as stuck-at-fault, I/V nonlinearities for high resistance states, and device-to-device variation in active conductance range, we found these have negligible impact on the final performance. }With this tool, we are able to analyze scaling up to more complex real-world problems.

    \section*{Scaled-up MANN for Mini-ImageNet}
    The methodology of crossbar-based TLSH and TCAM can be applied in many fields of deep learning that require the distance calculation and attention mechanism. 
    To show the scalability of our fully crossbar-based MANN, we conducted simulations based on our experimentally-calibrated model for one-shot learning using the Mini-ImageNet dataset\cite{vinyals2016matching}.  
    This dataset is derived from the ImageNet dataset with 100 classes of 600 images of size $84\times84$ color pixels per class. 
    The task is known to be much more difficult than that of the Omniglot handwritten dataset. 
    A more sophisticated ResNet-18 model is used as the controller following the state-of-the-art structure in few-shot learning models\cite{wang2019simpleshot}, which has more than 11 million weights, 44 times larger than the controller to classify images from the Omniglot dataset. 
    
    A challenge for this network is the required crossbar sizes (512 in one dimension), and thus the voltage drops along the wire would significantly reduce the computing accuracy. 
    % To enable this scaled-up MANN in crossbars for the Mini-ImageNet dataset, we consider the significant IR drop problem and gives an overview of the architecture for few-shot learning in Fig. \ref{fig:mini-imagenet}\textbf{a}. 
    % The CNN controller in crossbar systems has been widely reported\cite{wang2019situ,shafiee2016isaac,yao2020fully} which we don't elaborate it in this work. 
    % Since the dimension of output feature vectors from controller is 1,600 (see Methods), it is too large to implement TLSH and TCAM in a single array due to IR drop problem. 
    This is solved by partitioning large arrays (for hyperplanes and memories) into smaller 256$\times$256 tiled crossbars (Fig. \ref{fig:mini-imagenet}\textbf{a}) to accommodate the model. 
    % With the simulation on the Omniglot dataset as our backbone, we conduct a simulation with wire resistance on scaled-up MANN for the Mini-ImageNet dataset. 
    In the simulation, we consider the experimental device fluctuations (see Supplementary Table \ref{tb:fluc}) and use the same threshold current (\SI{4}{\micro\ampere}) for the TLSH approach as in the smaller Omniglot problem. 
    The result in Fig. \ref{fig:mini-imagenet}\textbf{b} shows that the classification accuracy for the 5-way 1-shot problem increases with the number of hashing bits, and reaches 58.7\% with 4,096 hashing bits, only 1.3\% smaller than the model implemented in digital hardware with cosine similarity as the distance metric. 
    We also explored the performance with different partitioned array sizes in Supplementary Fig. \ref{fig:mini_scale}\textbf{a, b}, which achieves nearly equivalent performance with arrays smaller than or equal to 256$\times$256 and drops slightly with the 512$\times$512 array. Encouragingly, the TLSH function can be implemented with a larger array (512$\times$512) because of lower conductance and smaller voltage drops along the wires. 
    % Fig. \ref{fig:mini-imagenet}\textbf{b} presents the simulation results of classification accuracy for the 5-way 5-shot problem. 
    % Due to significantly increasing complexity of this task, we can expect that we need a larger number of hyperplanes comparable to the size of inputs in order to achieve higher accuracy. 
    % Our model implemented in digital hardware has achieved 53.3\% classification accuracy on the 5way-5shot task with cosine distance as the distance metric.
    % As expected, the result shows that the classification accuracy on Mini-ImageNet increases with the increasing number of hashing bits, reaching 51.4\% at 4096 hashing bits with only 1.9\% degradation comparing to cosine similarity. 
    % We also explore the performance of TLSH and TCAM with different array size in Supplementary Fig. \ref{fig:mini_scale}\textbf{a, b}. 
    % We further find out that the array size for TLSH can be even extended to 512$\times$512 with the equivalent performance to the smaller size. 
    %However, we do discover the accuracy drop with increasing array size for TCAM because we apply larger conductance to the TCAM part. 
    From these results, we can see that the performance of our crossbar-based MANN can scale up effectively to at least Mini-ImageNet problems.

    \section*{Discussion}

\begin{figure}[htbp]
        \centering 
        \includegraphics[width=0.95\textwidth]{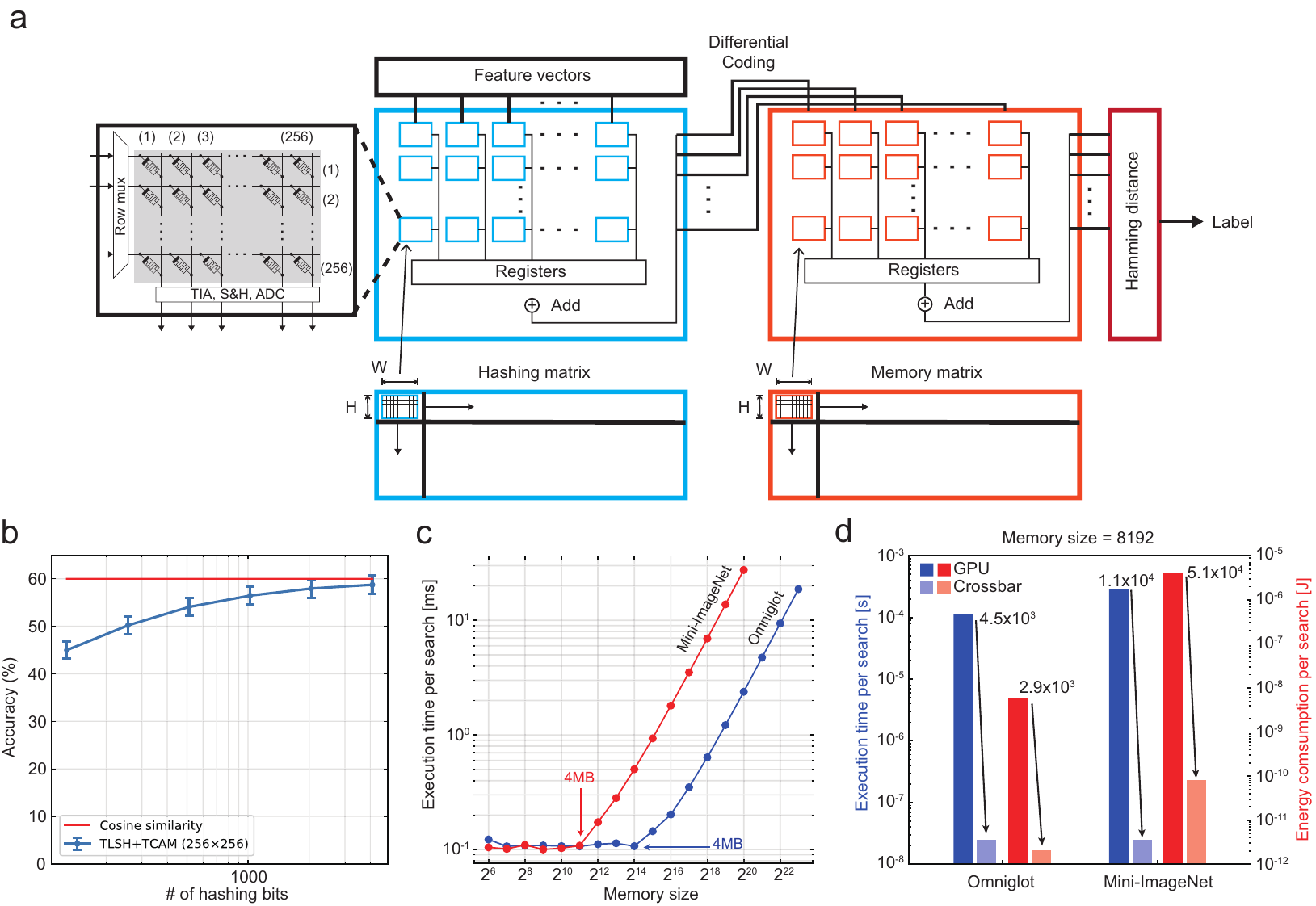}
        \caption{\textbf{Experiment validated simulation results of Mini-ImageNet dataset.} 
        \textbf{a}, The architecture of TLSH and TCAM for the scaled-up MANN. Both matrices for hashing and external memory are partitioned into tiled memristor crossbar arrays (H$\times$W) to mitigate the voltage drop problem in large crossbars and to increase the utilization rate. 
        % accelerate the vector-matrix-multiplication (VMM). 
        \textbf{b}, The accuracy performance from our experiment-validated models on Mini-ImageNet dataset.  
        The error bar shows the 95\% confidence interval among 100 repeated inference experiments.
        \textbf{c}, The execution time of search operations per inference on a GPU drastically increases when external memory size reaches a threshold, confirming the operation is memory intensive. 
        \textbf{d}, \rev{The comparison of the search latency and energy consumption for 5-way 1-shot learning on both the Omniglot and Mini-ImageNet datasets. 
        For GPU, the models for both datasets stores the same number of entries (8,192), but Mini-ImageNet uses a larger memory capacity due to higher dimension (64 vs. 512) of feature vectors, leading to even better improvement on latency and energy efficiency. The number of hashing bits used in crossbar arrays is 128 and 4096 for Omniglot and Mini-ImageNet, respectively.
        }% The number of hashing bits is 128 and 4096 for Omniglot and Mini-ImageNet, respectively. 
        }
        \label{fig:mini-imagenet}
    \end{figure}

    Compared to conventional von Neumann based implementations, the key advantage of crossbar-based MANN is lower latency and higher energy efficiency through co-located computing and memory, energy-efficient analog operations, and intrinsic stochasticity. 
    % Though the external memory enables neural networks to learn from rare events, its powerful abilities cannot be fully exploited due to the memory wall existing in traditional von Neumann computer architectures. 
    % When dealing with a more complex problem with a larger size of memory, a large amount of time is wasted in data transfer between the memory and the processing unit (CPU or GPU). 
    To evaluate the strength of the approach, we run the same 5-way 1-shot problem with Omniglot and Mini-ImageNet datasets on a digital graphic processing unit (GPU) (Nvidia Tesla P100).
    The time required to classify a single image increases dramatically after the size of the MANN's external memory capacity reaches a certain threshold (only several MB) because of the repeated off-chip data movement (see Fig. \ref{fig:mini-imagenet}\textbf{c}). 
    This problem on conventional hardware has been the major bottleneck preventing the widespread adoption of few-shot learning.
    The approach of directly computing in the memory, or crossbar, provides a plausible solution to address this bottleneck. 
    % However, with the in-memory computing characteristics of memristor crossbar arrays, we can avoid the memory transfer during learning, thus dramatically enhancing the performance of MANNs in terms of latency and energy and feasibility to adapt to more complex problems. 
    In the crossbar-based MANNs, the matrix multiplication in the convolutional layer, the hashing in TLSH, and the searching operation in TCAM are all computed with single-step current readout operations.
    With our current proof-of-concept experimental system, readouts take about \SI{100}{\nano\second}, but can be reduced to \SI{10}{\nano\second}. 
    We also considered the time latency for the peripherals (\SI{2.5}{\nano\second}) including the full-adders for tiled crossbar arrays\cite{ghadiry2013dlpa}.
    With these conservative forecasts, we compared the latency for GPU and our analog in-memory hardware. 
    The results shown in Fig. \ref{fig:mini-imagenet}\textbf{d} indicate a latency improvement (4,540$\times$ and 11,280$\times$) when the memory size (number of entries) is 8,192, for both Omniglot and Mini-ImageNet datasets. 
    \rev{Additionally, our approach also offers high energy efficiency compared with the conventional GPU (2,857$\times$ and 50,970$\times$). Detailed analysis about energy and latency estimation can be found in Supplementary Note \ref{sinote:energy}.}

    % \section*{Conclusion}
    In summary, we have, for the first time, experimentally implemented a complete MANN architecture, from the controller to distance calculation, in an analog in-memory platform with proven high robustness and scalability. 
    We utilize the analog behavior of memristor devices to perform convolution operations for CNNs and exploit the inherent stochasticity of devices to perform hashing functions. 
    A novel hardware-friendly hashing function (TLSH) is developed to provide better analog computing error tolerance and lower power consumption. 
    In addition, a differential encoding method for crossbar-based TCAM is applied to adapt to the ternary Hamming distance calculations requirements. 
    In our experiment, all dot-product operations are performed in physical crossbars, which exhibit experimental imperfections, such as device state fluctuations, device nonlinearities, voltage drops due to wire resistance, and peripheral circuits.  
    The fully hardware-implemented MANNs delivered similar accuracy compared to software on few-shot learning with the widely used the Omniglot dataset. 
    Simulation results on Mini-ImageNet show the ability of crossbar-based MANN to execute real-world tasks, with much-improved latency and energy consumption. 
    We demonstrate that analog in-memory computing with memristive crossbars efficiently supports many different tasks, including convolution, hashing, and content-based searching. The successful demonstration of these functions opens possibilities with other machine learning algorithms, such as attention-based algorithms, or reaching scales that are currently prohibited by conventional hardware (e.g, Fig. \ref{fig:mini-imagenet}\textbf{c}). Additionally, there are many opportunities for future software-hardware co-optimization to improve the accuracy and efficiency results further. 
    
    \begin{methods}
    
    \subsection{Memristor integration} 
    The memristors are monolithically integrated on CMOS fabricated in a commercial foundry in a \SI{180}{\nano\meter} technology node. 
    The integration starts with the removal of native oxide on the surface metal with reactive ion etching (RIE) and a buffered oxide etch (BOE) dip. 
    Chromium and platinum are then sputtered and patterned with e-beam lithography as the bottom electrode, followed by reactively sputtered \SI{2}{\nano\meter} tantalum oxide as the switching layer and sputtered tantalum metal as the top electrode. 
    The device stack is finalized by sputtered platinum for passivation and improved electrical conduction.

    \subsection{Iterative write-and-verify programming method}
    In this work, we use the iterative write-and-verify method to program memristor devices to the target conductance value. 
    First, we set a target conductance matrix and the corresponding tolerant programming error range. 
    After that, successive SET and RESET pulses are applied to the target devices followed by conductance readout with READ pulses. 
    If the device conductance is below the target conductance minus the tolerant error, a SET pulse is applied. 
    A RESET pulse is applied for conductance above the tolerant values, while the device has been programmed within the tolerant values are skipped to pertain the state. 
    For the crossbar-based MANN in this work, we apply the write-and-verify method to map the weights of the CNN controller and memories in the TCAM structure. During the programming process, we gradually increase the programming voltage and gate voltage as shown in Supplementary Table \ref{tb:program}. 
    The pulse width for both the SET and RESET process is \SI{1}{\micro\second}. 
    The tolerant range we set is \SI{5}{\micro\siemens} above or below the target conductance value. 
    
    \subsection{Adjacent Column Method}
    We apply the Adjacent Column Method (ACM)\cite{kazemi2020device} to map the conductance of memristors in crossbar arrays to weights in hashing planes. ACM subtracts the neighboring columns as shown in Fig. \ref{fig:TLSH}\textbf{b} to generate the hash codes. Hence, for a crossbar array with N+1 columns, the output of differential encoding contains N values which immensely saves the area. The mathematical representation is as follows: Provided that we get a random conductance map after programming which is:
    \begin{equation}
    \label{eq:Gmap}
    G_{map}=
    \begin{pmatrix}
    G_{1,1}&G_{1,2}&\cdots&G_{1,N}\\
    G_{2,1}&G_{2,2}&\cdots&G_{2,N}\\
    \vdots&\vdots&\ddots&\vdots\\
    G_{N,1}&G_{N,2}&\cdots&G_{N,N}
    \end{pmatrix}
    \end{equation}
    then the ACM method is equivalent to multiplying $G_{map}$ by a transformation matrix:
    \begin{equation}
    \label{eq:ACM}
    G_{hash}=G_{map}\times
    \begin{pmatrix}
    1&0&\cdots&0&0\\
    -1&1&\cdots&0&0\\
    0&-1&\cdots&0&0\\
    \vdots&\vdots&\ddots&\vdots&\vdots\\
    0&0&\cdots&-1&1\\
    0&0&\cdots&0&-1
    \end{pmatrix}
    =
    \begin{pmatrix}
    G_{1,1}-G_{1,2}&G_{1,2}-G_{1,3}&\cdots&G_{1,N-1}-G_{1,N}\\
    G_{2,1}-G_{2,2}&G_{2,2}-G_{2,3}&\cdots&G_{2,N-1}-G_{2,N}\\
    \vdots&\vdots&\ddots&\vdots\\
    G_{N,1}-G_{N,2}&G_{N,2}-G_{N,3}&\cdots&G_{N,N-1}-G_{N,N}
    \end{pmatrix}
    \end{equation}

    \subsection{Memristor model and simulations}
    We build a memristor model to simulate the conductance fluctuation, which is the most dominant non-ideality of our crossbar-based MANN. The behavior of conductance fluctuation is assumed to be a Gaussian nature which is as follows:
    \begin{equation}
    \label{eq:G_fluc1}
    G = G_0+\sigma\cdot\mathcal{N}(0, 1) 
    \end{equation}
    where $G_0$ is the conductance after programming, $\sigma$ describes the standard deviation of the fluctuation range. In the simulation we assume that the device only fluctuates for different VMM processes since in the real experiment the execution time of one VMM is very small (\SI{10}{\nano\second}) which is negligible compared to time between succesive input vectors (\SI{1}{\micro\second}). 
    After considering the device-to-device variations and fitting the parameters (see Supplementary Table \ref{tb:fluc}) to experimental measurements (see Supplementary Fig. \ref{fig:fluc}), Equation \ref{eq:G_fluc1} becomes:
    \begin{equation}
    \label{eq:G_fluc2}
    G = G_0+\exp(a\cdot\ln(G_0)+b+s\cdot\mathcal{N}(0, 1))\cdot\mathcal{N}(0, 1)
    \end{equation}
    with $\mathcal{N}(\mu, \sigma^2)$ being the normal distribution with mean $\mu$ and standard deviation $\sigma$. In the simulation, we also consider the program error to the initial conductance $G_0$ which is shown as:
    \begin{equation*}
    \label{eq:programerr}
    G_0 = G_t + \mathcal{N}(0, \tilde{G_{err}^2})
    \end{equation*}
    where $G_t$ is the target conductance we want to program to and $\tilde{G_{err}}$ is the program error which we set to \SI{5}{\micro\siemens} in the simulation. 
    
    To get the parameters of the memristor model in terms of the effect of device fluctuation, we \emph{SET} 4,096 devices to 16 distinct analog states and \emph{READ} each device for 1,000 times. The relation between the mean value and standard deviation of 1,000 reads is shown in Supplementary Fig. \ref{fig:fluc}\textbf{a} and \ref{fig:fluc}\textbf{b}. 
    We further analyze the standard deviation distribution for each conductance state from \SI{5}{\micro\siemens} to \SI{50}{\micro\siemens}, plot the distributions in logarithmic scale, and fit them with Gaussian distribution. 
    The results are shown in Supplementary Fig. \ref{fig:fluc}\textbf{c}. The mean value of $s$ for each distribution gives us the parameter for the model. In addition, we fit a linear curve with conductance states and standard deviation in a log-log regime of measurements (see Supplementary Fig. \ref{fig:fluc}\textbf{d}). The fitted parameters $a$ and $b$ are used in the simulation.

    \subsection{Ternary locality sensitive hashing}
    Ternary locality-sensitive hashing introduces a wildcard 'X' to the hashing vector to alleviate the analog computing error from nonideal factors. 
    We have demonstrated that this modified hashing scheme can achieve software-equivalent performance (LSH with the same hashing bits) on our crossbar arrays. 
    The threshold current $I_\text{th}$ applied in the experiment should be carefully chosen according to the typical value of the computing error caused by device fluctuation. 
    The value we chose throughout the experiment is \SI{4}{\micro\ampere}. 
    We also show the dependence of classification accuracy on different threshold currents in Supplementary Fig. \ref{fig:tlsh_x}. 
    
    For the simulation results in Fig. \ref{fig:CNN}\textbf{d} and \ref{fig:CNN}\textbf{e}, where the device fluctuation varies, 
    % since we increase the device fluctuation level to find the tolerant range, 
    we chose different threshold currents $I_\text{th}$ according to the fluctuation levels. 
    Specifically, for our memristor model which can be described by Equation \ref{eq:G_fluc1}, we empirically set the threshold current to be $5\sigma\cdot V_{in}$ where $V_{in}$ is the maximum input voltage to the row line when performing VMM. The $V_{in}$ is chosen to be \SI{0.2}{\volt} in our experiment.

    To generate random hashing planes in crossbar arrays (Fig. \ref{fig:TLSH}\textbf{c}), we \emph{RESET} the devices from an arbitrary high conductance state to near \SI{0}{\micro\siemens}, where the conductance is ultimately decided by the intrinsic stochastic switching process. Regardless of the initial states, we use 5 \emph{RESET} pulses with an amplitude of \SI{1.5}{\volt} and a width of \SI{20}{\nano\second}. The \emph{RESET} voltage is carefully controlled to protect memristor devices because larger voltages may cause devices to be stuck at low conductance states.

    \subsection{CNN architecture}
    The convolutional neural network (CNN)s in crossbars is applied as the controller in the MANN to extract features from incoming images. 
    
    The CNN structure for the Omniglot dataset is composed of:
    \begin{itemize}
        \item 2 convolutional layers, each with 32 channels of shape 3x3
        \item A 2x2 max-pooling layer
        \item 2 convolutional layers, each with 64 channels of shape 3x3
        \item A 2x2 max-pooling layer
        \item A fully connected layer with 64 outputs
    \end{itemize}
    Each convolutional layer is followed by a rectified linear unit (ReLU) activation layer. 
    
    The ResNet-18 for the Mini-ImageNet dataset is composed of 8 residual blocks. Each residual block has two 3x3 convolutional layers with size \textit{[layer channel, input channel, 3, 3]} and \textit{[layer channel, layer channel, 3, 3]} respectively. Each convolutional layer is followed by a batch normalization layer and a ReLU layer. The overall architecture for ResNet-18 is:
    \begin{itemize}
        \item 1 convolutional layer with 64 channels of shape 3x3
        \item 2 residual blocks with 64 channels and stride 1
        \item 2 residual blocks with 128 channels and stride 2
        \item 2 residual blocks with 256 channels and stride 2
        \item 2 residual blocks with 512 channels and stride 2
        \item 1x1 adaptive average pooling layer
    \end{itemize}

    We map the weights of \rev{convolitional layers} in the CNN to the conductance of memristor devices using differential encoding\cite{yao2020fully}. To elaborate, in a differential column pair, we program the positive weight to the left column and the absolute value of negative weights to the right column, while keeping the other at the low conductance state. The weight-to-conductance ratio we set in our experiment is 1:\SI{50}{\micro\siemens}. 
    The feature maps collected from the output current in Fig. \ref{fig:CNN}\textbf{a} are converted into voltages and then sent to another crossbar array corresponding to the subsequent convolutional layer. 
    \rev{The fully connected layer is retrained after mapping convolutional layers on crossbar arrays and it's computed in the digital domain.}

    \subsection{Memory update rules}
    In an $N$-way $K$-shot learning, the memory module is updated based on the $N \times K$ images in the support set. 
    If the label of the new input image label doesn't match the label of the nearest neighbor entry, we simply find a new location in the memory and write the input image to that location. Conversely, if the input image label matches, we need to update the memory of the nearest neighbor. In the GPU, we assign cosine distance as the metric to identify the label of input images and update the real-valued vectors at the same location\cite{kaiser2017learning}. However, in this work, we use ternary Hamming distance as our metric, and we apply the following rules to update the ternary vectors in the external memory: We introduce a scoring vector to evaluate the majority of '1' and '0' for each bit of each memory vector. An element-wise mapping function is applied to each ternary vector stored in the memory module:
    \begin{equation*}
    \label{eq:mapping}
    f: \{1, 0, X\}^D \rightarrow \{1, -1, 0\}^D
    \end{equation*}
    where $D$ is the dimension of storing vectors. For example, vector $(1, 0, 1, X, 0)$ is mapped to $(1, -1, 1, 0, -1)$. We assume the scoring vector for each storing vector as: $s_i = f(a_i), i = 1, 2, 3, \ldots, M$, where $s_i$ is the scoring vector, $a_i$ is the hashing vector stored in TCAM and $M$ is the total number of memory entries. When there is a match case happening at memory location $k$, we first update its scoring vector as below:
    \begin{equation}
    \label{eq:score_update}
    s_k^{*} = s_k \oplus  f(v) 
    \end{equation}
    where $s_k^{*}$ is the new scoring vector, $v$ is the hashing vector of the new image, $\oplus$ is element-wise add operation. Then we update the memory at the same location using the following rules:
    \begin{equation}
    \label{eq:mem_update}
    a_i^{*} = L(a_i); L(x)=
    \left\{
    \begin{array}{ll}
        1 &  x>0 \\
        X &  x=0 \\
        0 &  x<0
    \end{array}
    \right.
    \end{equation}
    where $a_i^{*}$ is the updated memory and $L$ is an element-wise operator. Therefore, bits of the memory stored in TCAM are decided by the majority of '1' and '0' of incoming vectors which match the storing vectors.

    \subsection{Omniglot training}
    The Omniglot data set contains 1623 different handwritten characters from 50 different alphabets. 
    Each character was drawn online by 20 different people using Amazon's Mechanical Turk. 
    In the experiment, we augment the 964 different characters in the training set to 3856 through rotation. 
    The character types in the test set remain unchanged at 659. 
    There is no overlap between the training set and the test set. We use the episode training method during the training process. 
    Episode training is to select N x M instances from the training set during each training, where N represents different classes and M represents the number of instances in each class. 
    The purpose of episode training is to enable the learned model to focus on the common parts, ignoring tasks, so as to achieve the purpose of learning to learn.
    The specific settings in the training process are as follows: memory size is 2048, batch size is 16, episode width is 5, episode length is 30; The length of the output key is 128, and the validation set is used for verification every 20 times.

    \subsection{Mini-ImageNet training}
    The Mini-ImageNet dataset contains 100 classes randomly chosen from the ImageNet dataset. We randomly split the dataset into a training set, a validation set, and a test set containing 64, 16, and 20 classes, respectively. 
    We take the pre-trained model in ref\cite{wang2019simpleshot} and fine-tuned it using cosine distance as the meta-training metric. Once the meta-training process is done, the weights for the controller won't be updated. 
    We use the ResNet-18 model as the CNN controller and the output feature vector of the CNN is 512-dimensional.

    \end{methods}
    
    \section*{Data availability}
    The data supporting plots within this paper and other findings of this study are available with reasonable requests made to the corresponding author.

    \section*{Code availability}
    The code used to generate the results of this study is available with reasonable requests made to the corresponding author.

    \clearpage
    \section*{References}
    \:
    \bibliographystyle{naturemag}
    \bibliography{main}

    \begin{addendum}
     \item[Acknowledgement] 
   
     \item[Author contribution] C.L., C.G., J.P.S contributed to the conception of the idea. 
     R.M. performed the experiments and analyzed data under the supervision of C.L..
    R.M., Y.Z. and A.K. performed simulations. X.S. integrated the memristors. 
     R.M., Y.H, and C.L. wrote the manuscript with input from all authors.
   
     \item[Competing Interests] The authors declare that they have no competing interests.
     \item[Correspondence] Correspondence and requests for materials
    should be addressed to canl@hku.hk
    \end{addendum}

    % Start supplementary
    \beginsupplement

    % \@title
    % \section*{Supplementary Figures}
    \section{Supplementary Figures}

    \begin{figure}[!hth]
        \centering 
        \includegraphics[width=0.95\textwidth]{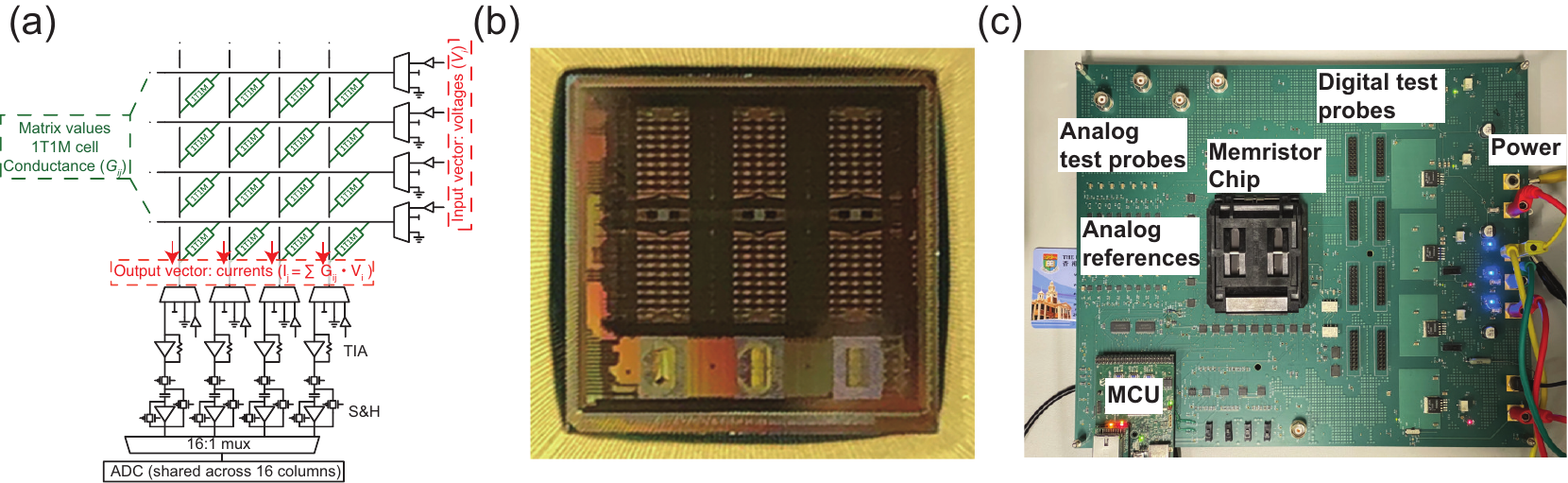}
        \caption{\rev{\textbf{CMOS-integrated crossbar test platform and experimental setup.}
        \textbf{a}, Circuit schematic of a crossbar for matrix multiplication with peripheral circuits. It consists of a 1T1R crossbar, row muxes, column muxes, transimpedence amplifiers (TIA), sample and hold (S\&H) and ADC.
        \textbf{b}, Picture of a wire-bonded integrated memristor chip. It constains three 64$\times$64 1T1R crossbar arrays.
        \textbf{c}, Measurement board with the integrated memristor chip that connects to a general purpose computer through a microcontroller (MCU).
        }
        }
        \label{fig:setup}
    \end{figure}

    \begin{figure}[!hth]
        \centering 
        \includegraphics[width=0.95\textwidth]{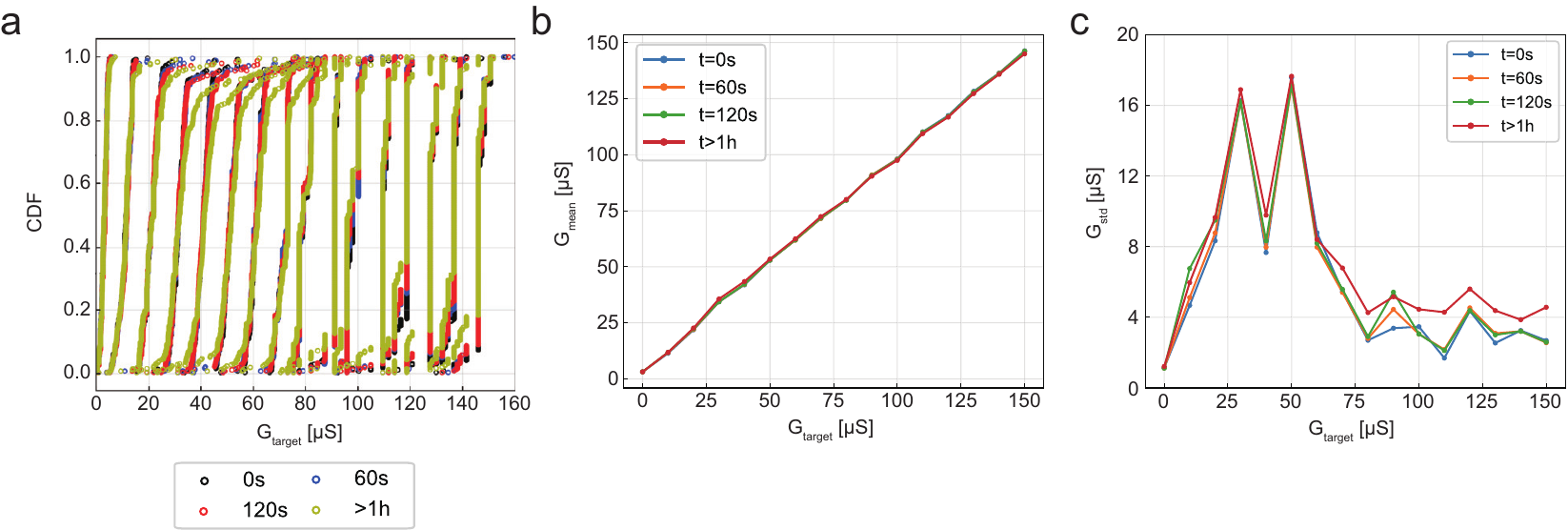}
        \caption{\textbf{Memristor conductance relaxation statistics in a crossbar array.} 
        \textbf{a}, Cumulative distribution function on 16 distinct programmed conductance states at different time period. We run the test on a 64x64 array where we divide it into 4x4 blocks and program them to 16 conductance states with iterative write-and-verify programming scheme. 
        \textbf{b}, Relation between mean conductance at different time periods and target conductance.  
        \textbf{c}, Relation between the standard deviation at different time periods and the target conductance. }
        \label{fig:Grelax}
    \end{figure}
    
    \begin{figure}[!hth]
        \centering 
        \includegraphics[width=0.95\textwidth]{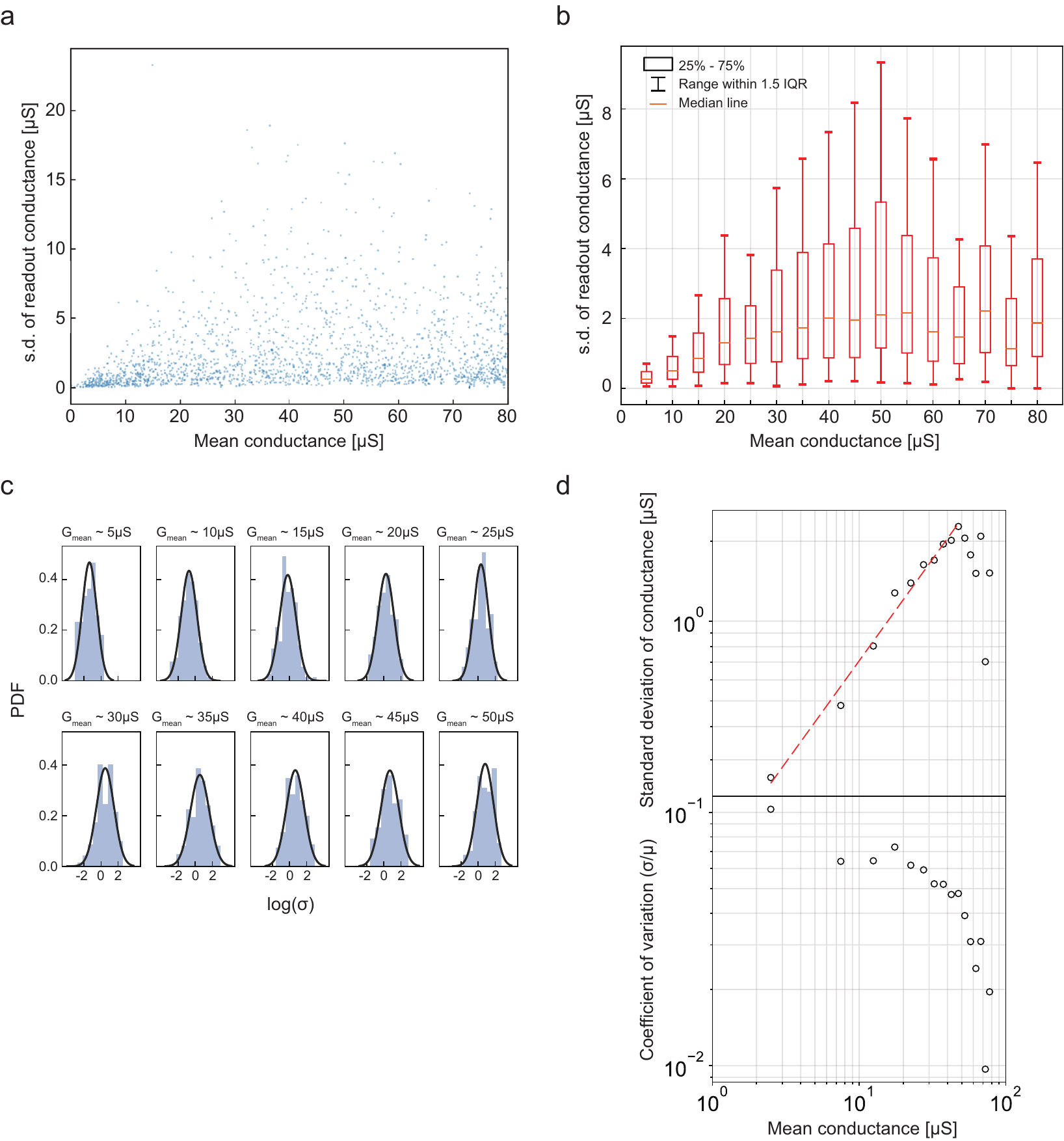}
        \caption{\textbf{Memristor conductance fluctuations  in a crossbar array.} 
        \textbf{a}, Relation between the standard deviation of 1000 readout conductance and the mean value at different conductance states. We test the whole 64x64 array. 
        \textbf{b}, Boxplot of the standard deviation with respect to different conductance states. 
        \textbf{c}, Distribution of logarithm of standard deviation at different conductance levels. The distributions show that the device-to-device variation in terms of conductance fluctuation exhibits a lognormal distribution.
        \textbf{d}, Mean value of the logarithm of standard deviation exhibits a linear dependence on the conductance state at low conductance range ($<$ \SI{50}{\micro\siemens}) and tends to maintain within a certain range at higher conductance range. \rev{The coefficient of the variation brought by the read fluctuation decreases as the conductance increases.}
        }
        \label{fig:fluc}
    \end{figure}

    \begin{figure}[!hth]
        \centering 
        \includegraphics[width=0.95\textwidth]{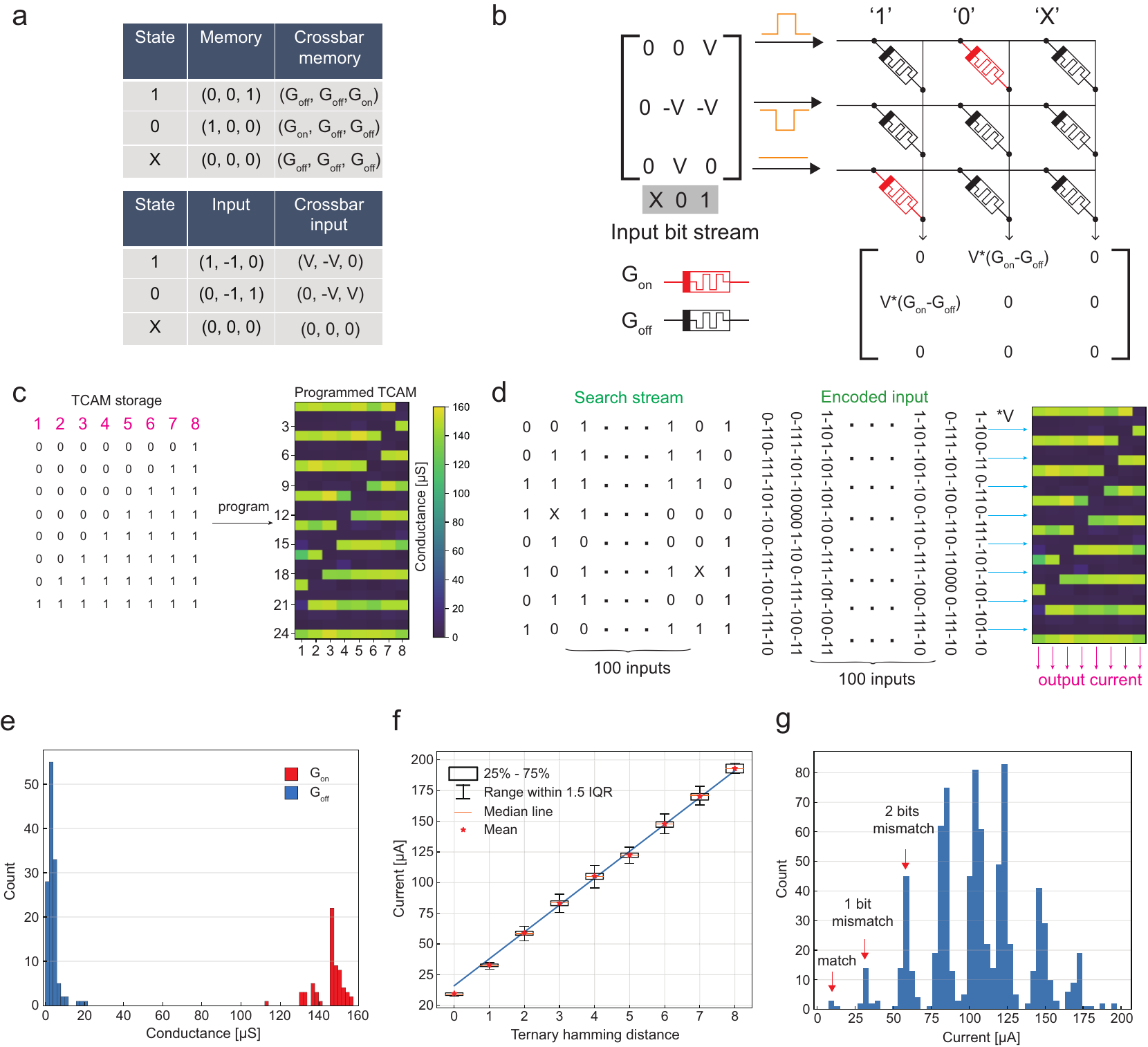}
        \caption{\textbf{3-bit encoding method for TCAM using crossbar array.} 
        \textbf{a}, TCAM storage and input encoding method of 3-bit encoding. 
        \textbf{b}, The schematic of experimental set-up for THD calculation by using dot product. Three bits are stored in crossbar array using 3-bit encoding. Search bits are successively applied to the row line and the output current from column line exhibits the THD between input and memory. 
        \textbf{c}, Eight binary vectors stored in crossbar as memory after 3-bit encoding. 
        \textbf{d}, 100 different ternary vectors sent to TCAM to perform THD calculation in a parallel way. 
        \textbf{e}, Distribution of G\sub{on} and G\sub{off}. 
        \textbf{f}, Ouput current exhibits a linear dependence on the THD. IQR, interquartile range. 
        \textbf{g}, Current distributions are separated from each other through which we can obtain the number of mismatch bits, \ie, THD. }
        \label{fig:3dpe}
    \end{figure}

    \begin{figure}[!hth]
        \centering 
        \includegraphics[width=0.95\textwidth]{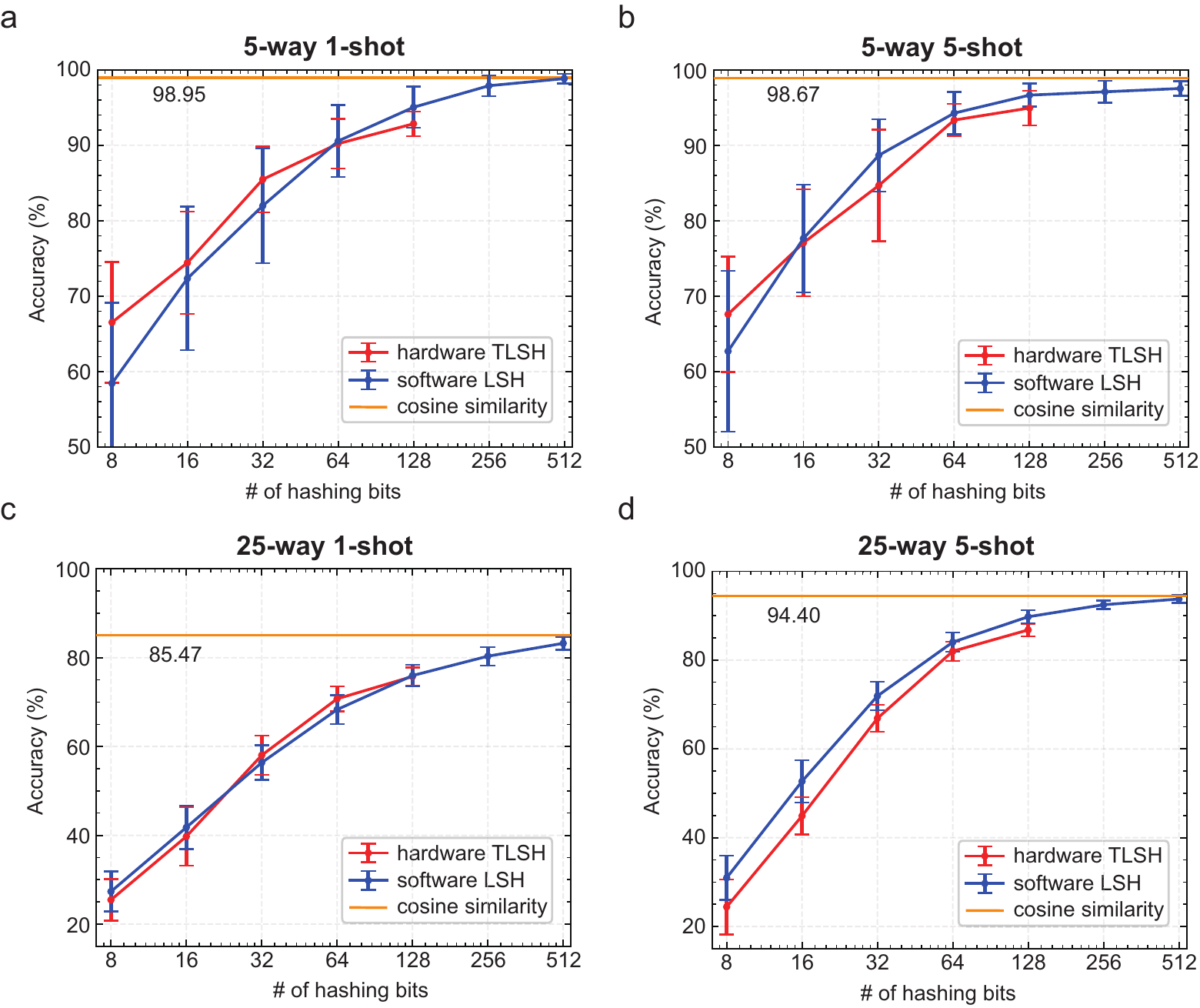}
        \caption{\textbf{Robustness of hardware TLSH on few-shot learning.} 
        \textbf{a, b, c, d}, To demonstrate the robustness of hardware TLSH which exploits the intrinsic stochasticity of memristors, we randomly choose 20 different conductance maps to perform the hashing process. The figure shows the classification accuracy with hardware TLSH and software LSH based on the experimentally generated hashing codes. The result exhibits the robustness of our memristor crossbar array served as hashing vector generator whose behavior is nearly the same as software LSH. Furthermore, we can obtain the same accuracy as cosine similarity on few-shot learning tasks if we scale the hashing bits up to 512 which is feasible in the prevailing crossbar architectures.
        }
        \label{fig:multi_lsh}
    \end{figure}

    \begin{figure}[!hth]
        \centering 
        \includegraphics[width=0.95\textwidth]{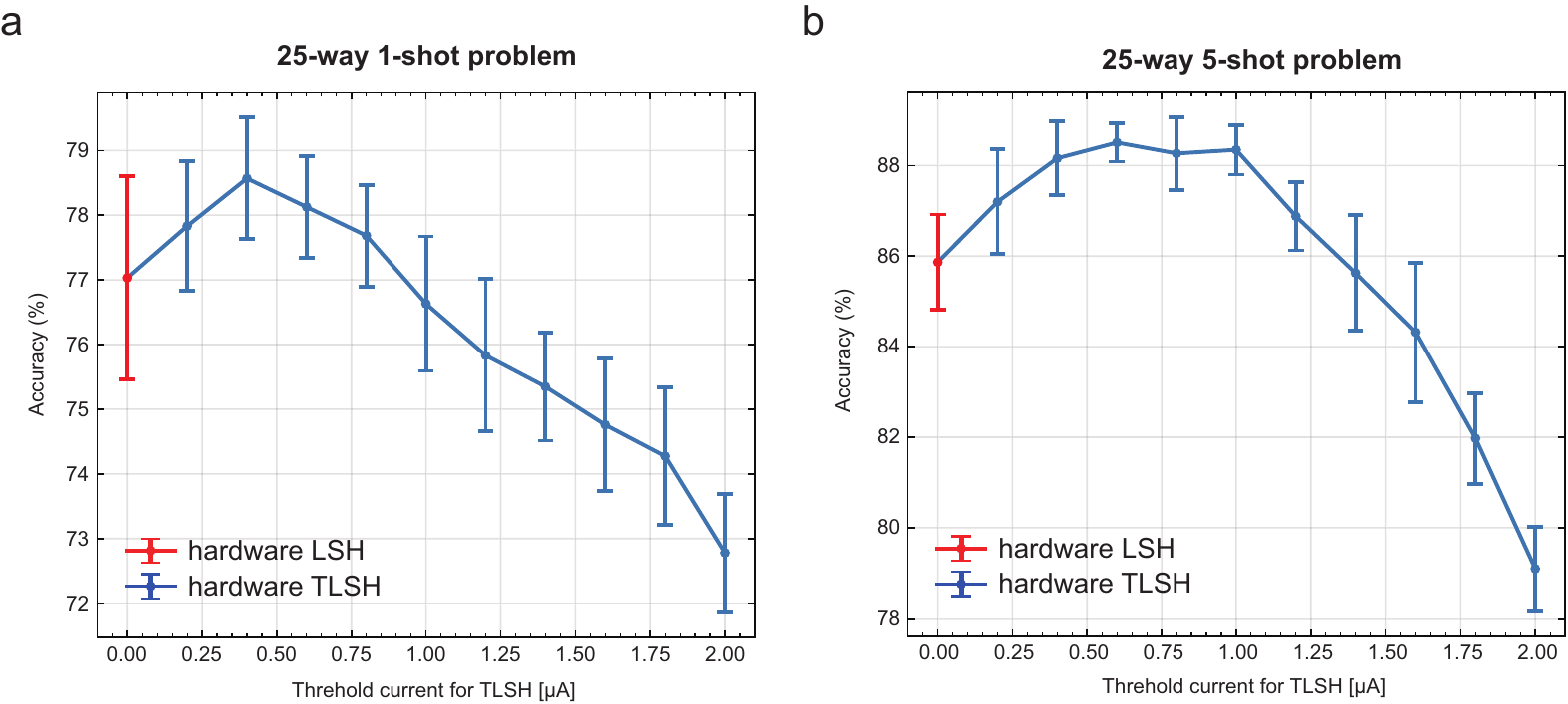}
        \caption{\textbf{Impact of the threshold current for hardware TLSH on the classification accuracy.} 
        \textbf{a, b}, As expected, the influence of the threshold current for hardware TLSH should be positive first and negative afterward. Because introducing the wildcard 'X' in LSH can mitigate the defect of device nonidealities but can also cast away the information of original input vectors. Here, we experimentally demonstrate the classification accuracy of 25-way problems with different threshold currents. The result shows that for our system, the performance reaches a peak with a threshold current around \SI{0.4}{\micro\ampere}. The zero thresholds current case denotes hardware LSH. Each task is repeated 100 times.
        }
        \label{fig:tlsh_x}
    \end{figure}

    \begin{figure}[!hth]
        \centering 
        \includegraphics[width=0.95\textwidth]{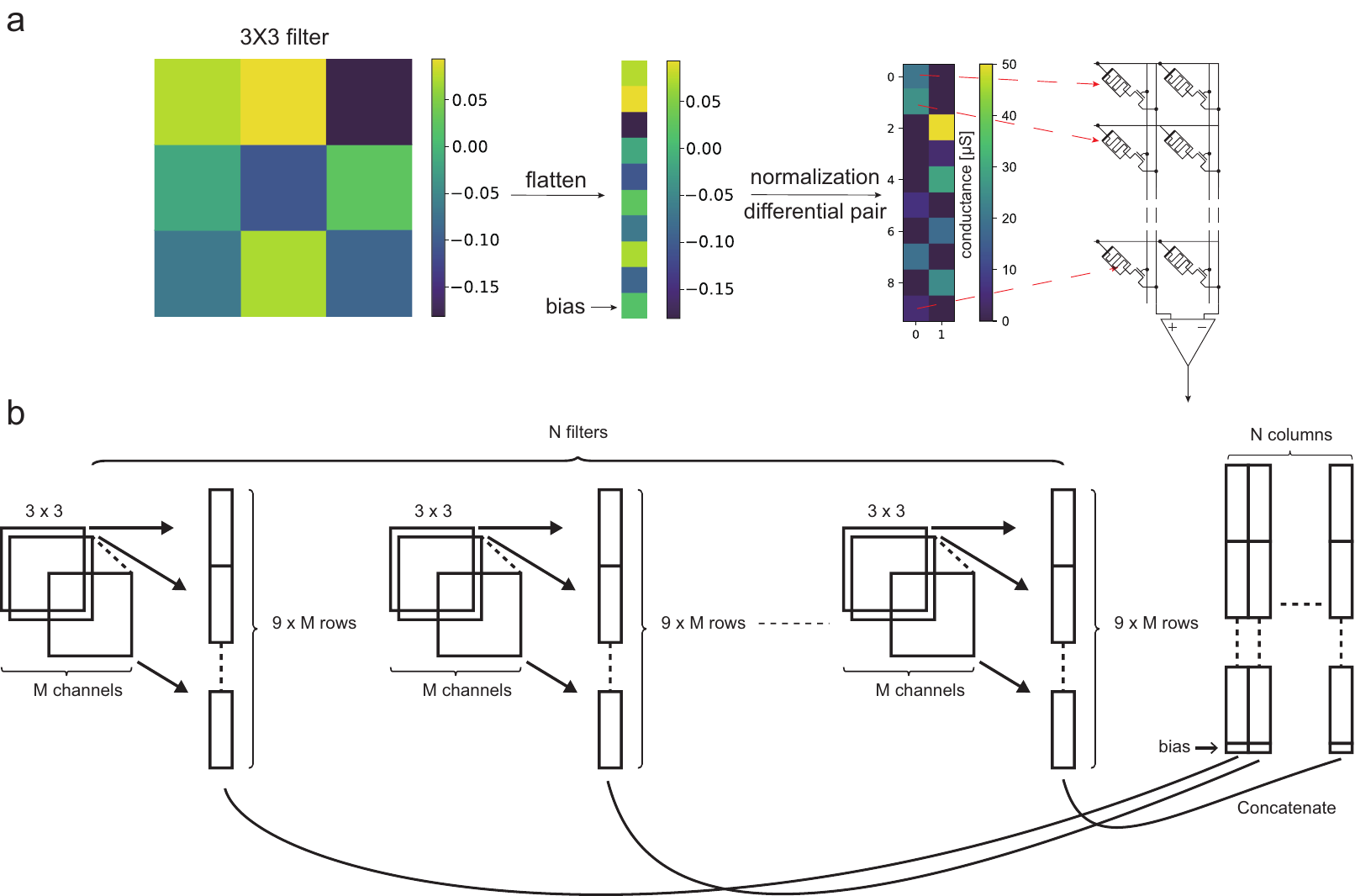}
        \caption{\textbf{Convolutional layers mapped to crossbar arrays} 
        \textbf{a}, A $3\times3$ convolutional kernel is flattened first and mapped to conductance using differential pair. 
        \textbf{b}, N filters with each containing M channels are concatenated together and mapped to crossbar arrays. Different filters perform the convolution in a parallel way.
        }  
        \label{fig:cnnlayer}
    \end{figure}

    \begin{figure}[!hth]
        \centering 
        \includegraphics[width=0.85\textwidth]{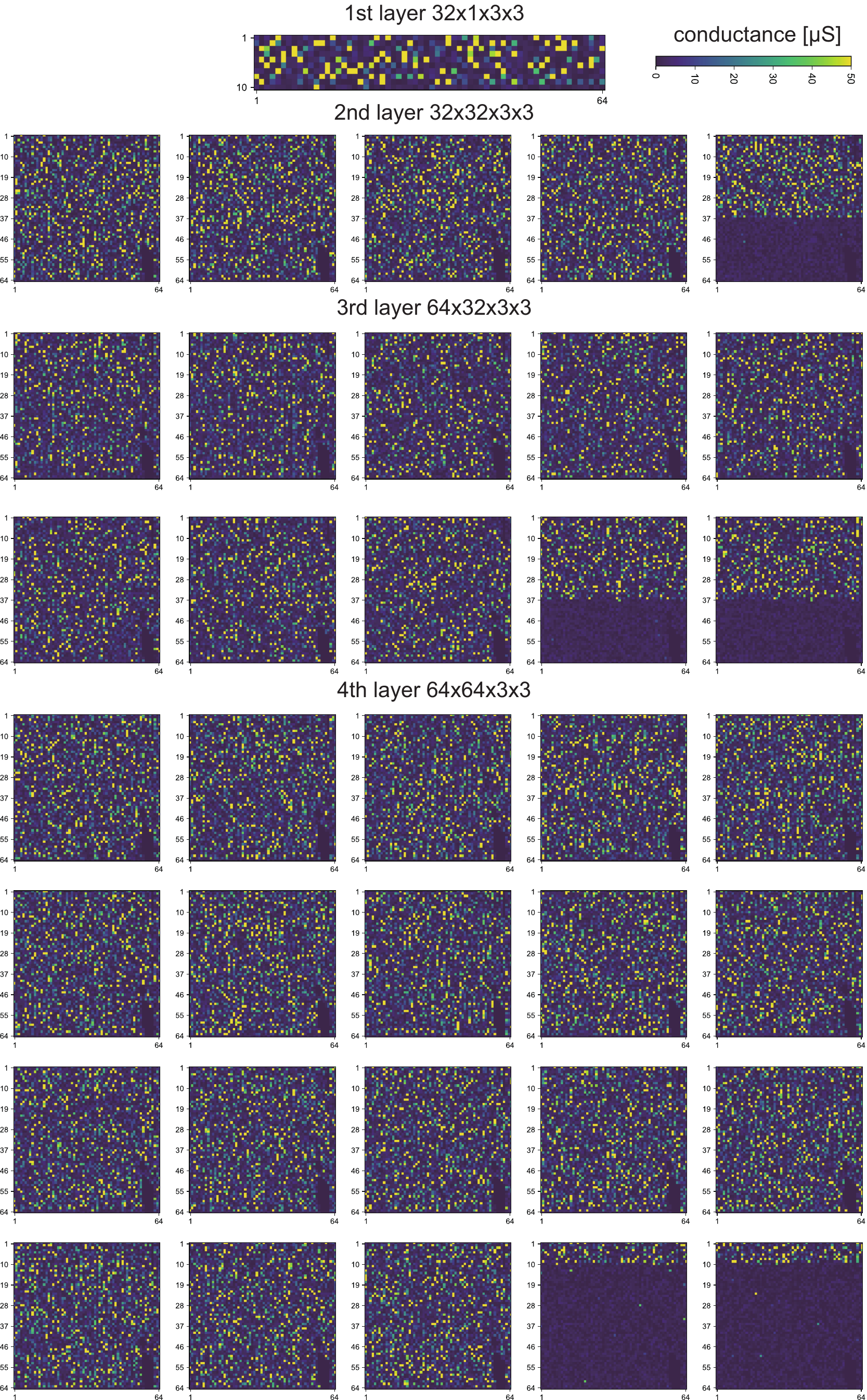}
        \caption{\textbf{Experimental conductance map of convolutional layers}
        }  
        \label{fig:cnngmap}
    \end{figure}

    \begin{figure}[!hth]
        \centering 
        \includegraphics[width=0.95\textwidth]{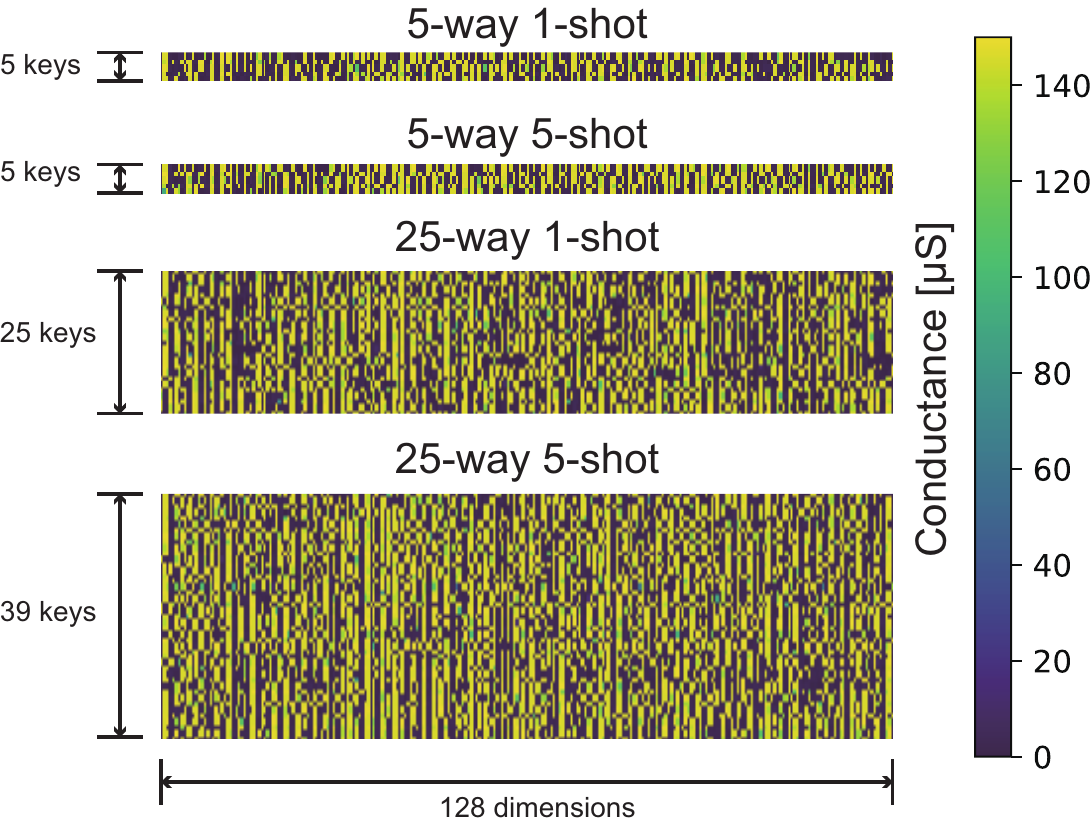}
        \caption{\rev{\textbf{Memory stored in the crossbar-based TCAM after TLSH operation and binary update process}
        }
        }  
        \label{fig:tcam_gmap}
    \end{figure}

    \rev{
    \begin{figure}[!hth]
        \centering 
        \includegraphics[width=0.95\textwidth]{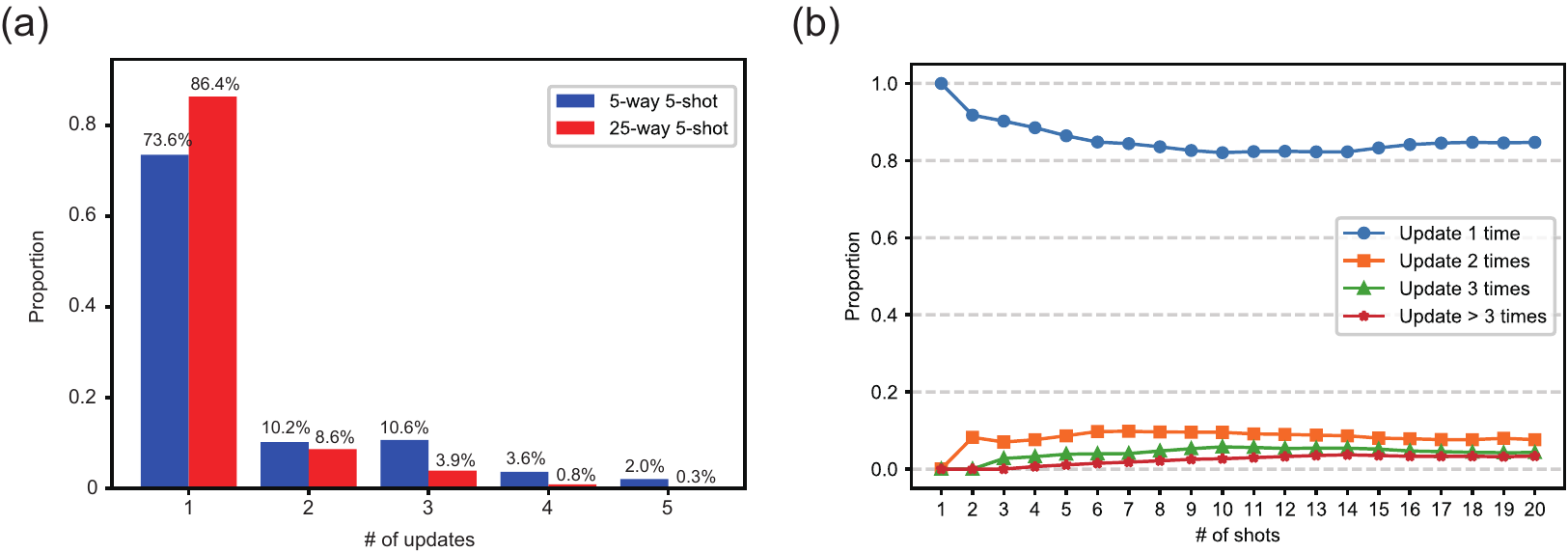}
        \caption{\rev{\textbf{Update times of the memory stored in crossbar-based TCAM of multi-shot learning.} 
        The majority of the bits are updated only once which is when writing the newly input vector into the new location of the memory. Unlike the conventional memory update method which need to update every value in the real-valued vectors, our proposed binary update method only updates very few bits in the memory which is suitable for the life-long learning given the endurance of memristors. 
        \textbf{a, } Statistical view of the update times of bits during standard 5-shot learning.
        \textbf{b, } To analyze the life-long learning properties, we perform the multi-shot learning on the 25-way task from 1-shot to 20-shot. It shows that nearly 80\% of the bits stored don't need to be updated once written to store the new input vector. There are only smaller than 5\% of bits that need to be updated higher than 3 times throughout the 20-shot learning.
        }
        }  
        \label{fig:binary_updates}
    \end{figure}
    }

    \begin{figure}[!hth]
        \centering 
        \includegraphics[width=0.95\textwidth]{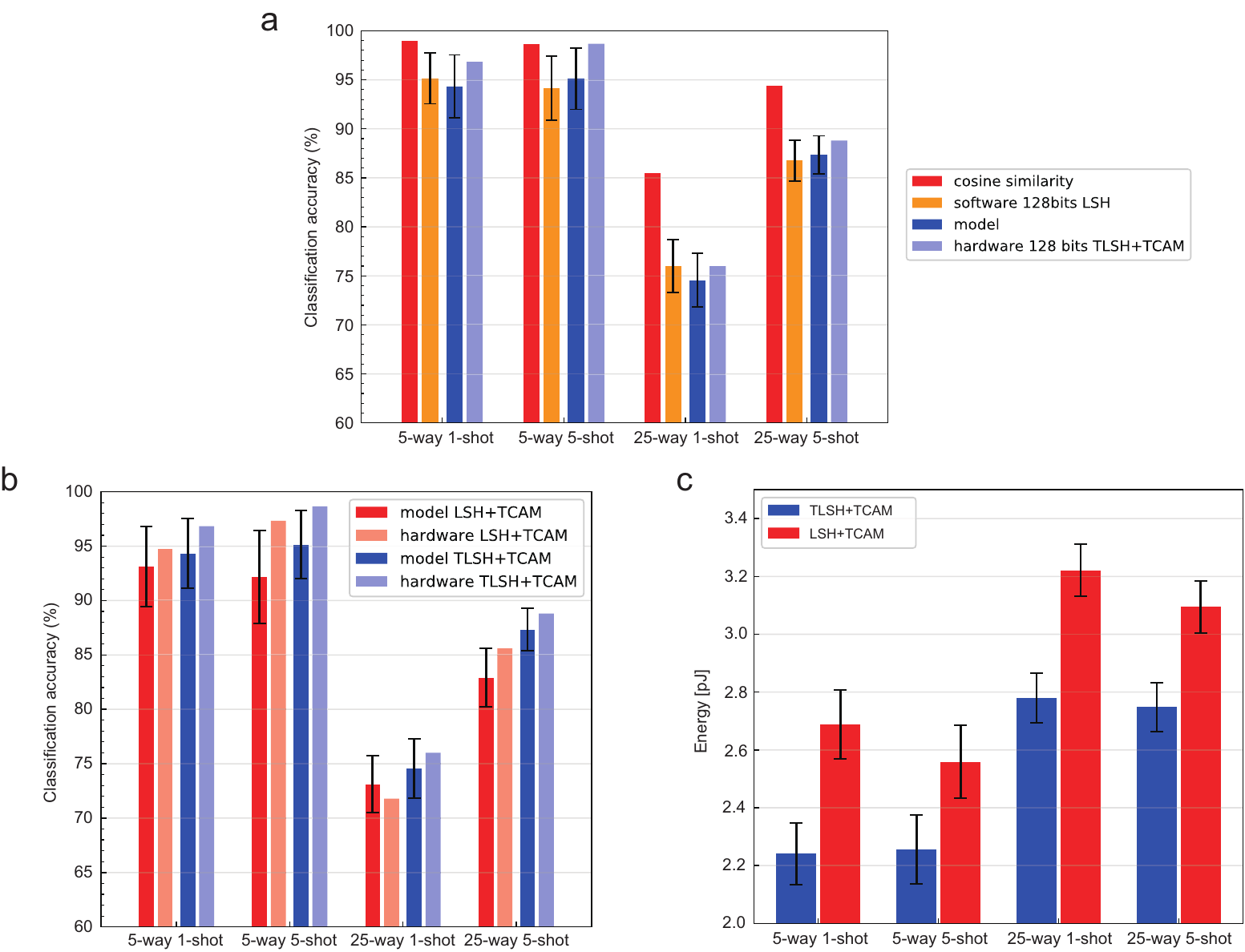}
        \caption{\textbf{Simulation results on Omniglot dataset accord with experimental results} 
        \textbf{a}, Classification accuracy of derived memristor device model, along with cosine similarity, software-based LSH with 128 bits, and end-to-end experimental results on crossbar arrays. 
        \textbf{b}, Comparison over LSH and TLSH with both simulation and experiment on few shot learning tasks.
        \textbf{c}, Average energy consumption of each TCAM search operation with TLSH and LSH, respectively. 
        }
        \label{fig:tlsh_lsh}
    \end{figure}

    \begin{figure}[!hth]
        \centering 
        \includegraphics[width=0.95\textwidth]{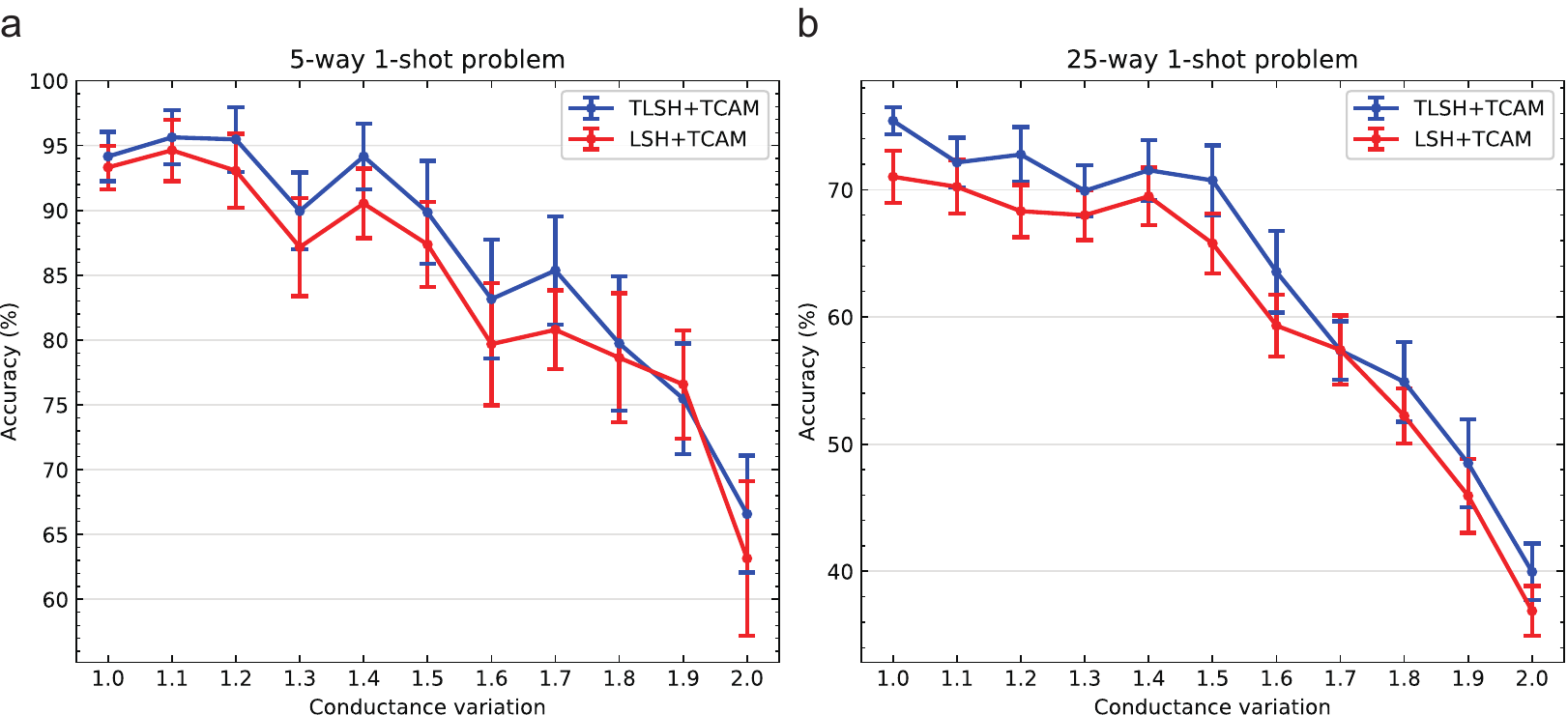}
        \caption{\textbf{Impact of device variation on classification accuracy.} 
        \textbf{a, b} The experimental device-to-device variation is shown in Supplementary Fig. \ref{fig:fluc}\textbf{c}. In simulation, we increase the variation to explore the influence of it on classification accuracy. The results show that for both 5-way 1-shot and 25-way 1-shot problem, the classification accuracy drops by 10\% when conductance variation is about 50\% larger than the experiemntal data. The value of conductance variation is the value of parameter $s$ in Supplementary Table \ref{tb:fluc}
        }
        \label{fig:acc_var}
    \end{figure}

    \begin{figure}[!hth]
        \centering 
        \includegraphics[width=0.95\textwidth]{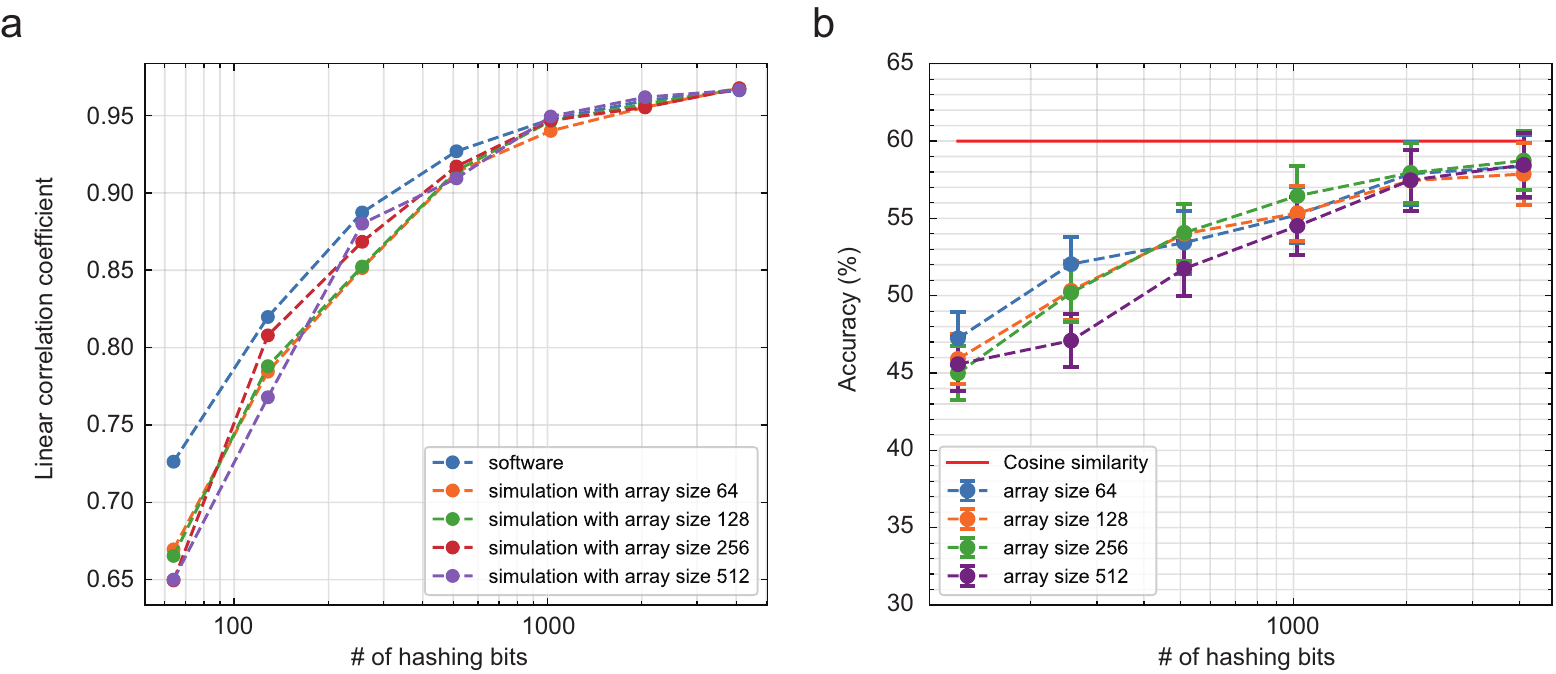}
        \caption{\textbf{Performance of scaled-up MANNs with different array size} 
        \textbf{a}, TLSH performance with different array size. The sneak path problem has no influence on the TLSH part because most of the devices are at the OFF state and the hyperplanes are random. 
        \textbf{b}, Classification accuracy with different array size. The performance degradation is found with a bigger array size and a small number of hashing bits. This is because the sneak path problem affects the performance of TCAM, resulting in miscalculating of Hamming distances. However, the accuracy remains nearly the same for different partitioned array sizes when the number of bits is large. 
        }
        \label{fig:mini_scale}
    \end{figure}

    \section{Supplementary Tables}

    \begin{table}[H]
        \centering
        \caption{Iterative write-and-verify method}
        \begin{tabular}{l|lll}
        \hline
        Program parameter & V\textsubscript{1}/V & V\textsubscript{2}/V& $\Delta$V/V \\
        \hline
        SET voltage & 1.0 & 2.5 & 0.1 \\
        RESET voltage & 0.5 & 3.5 & 0.05 \\
        SET gate voltage & 1.0 & 2.0 & 0.1 \\
        RESET gate voltage & 5.0 & 5.5 & 0.1 \\
        \hline
        \end{tabular}
        \label{tb:program}

    \end{table}

    \begin{table}[H]
        \centering
        \caption{Derived values of memristor model parameters}
        \begin{tabular}{l|ll}
        \hline
        Parameter & Description & Value\\
        \hline
        $G_0$ & Initial conductance after program & - \\
        $a$ & Linear fitting parameter & 0.782 \\
        $b$ & Linear fitting parameter & -2.168 \\
        $s$ & Device-to-device variation & 0.983 \\
        \hline
        \end{tabular}
        \label{tb:fluc}

    \end{table}
    %     \begin{table}[H]
    %         \centering
    
    %         \begin{tabular}{l|llll}
    %         \hline
    %         Operation & SL\_hi & SL\_lo   & DL1    & DL2    \\
    %         \hline
    %         Set M1    & Vset & 0      & Vg,set & 0      \\
    %         Reset M1  & 0    & Vreset & V\textsubscript{DD}     & 0     \\
    %         Set M2    & Vset & 0      & 0      & Vg,set \\
    %         Reset M2  & 0    & Vreset & 0      & V\textsubscript{DD}    \\
    %         Read M1   & Vread & 0     & V\textsubscript{DD}     & 0     \\
    %         Read M2   & Vread & 0     & 0      &V\textsubscript{DD}     \\
    %         \hline
    %         \end{tabular}
    %         \caption{ Write operation of the analog CAM cell}
    %         \label{tb:si_write}
    %     \end{table}
    % \pagebreak

    \pagebreak
    \section{Supplementary Notes}
    
    \rev{
    \begin{sinote}
        \textbf{Sensing margin in crossbar-based TCAM}
        % \can{did you discuss the `max sensing bits'?}

        % Here we analyze the sensing margin ()
        The sensing margin between the match and mismatch cases in conventional TCAM measures the reliability of the hardware under extreme conditions. 
        For TCAM circuits based on emerging memory devices, a higher sensing margin also provides better tolerance to device variation and thus a smaller bit error rate. 
        
        In this work, our crossbar-based TCAM does not only distinguish between match and mismatch cases but also returns the number of mismatches. 
        For few-shot learning with MANN, what is required is distinguishing between the closest match and the next close match. 
        Here we describe the sensing margin $\beta$ for our crossbar-based TCAM in detecting the closest match in Equation \ref{eq:sieq1}.

        % The sensing margin and the number of mismatch bits that can be detected are two important factors of TCAM. It often shows how reliable the TCAM is under extreme conditions. Especially for TCAM circuits based on emerging memory devices with non-negligible variations, a higher sensing margin tends to have better performance. In this work, our proposed TLSH plus TCAM structure can offer a high sensing margin for distance search applications. Here we perform a detailed analysis of the strength and limitations of our TLSH plus TCAM.

        % \subsection{Theoretical analysis}
        % Sensing margin and the number of mismatch bits that can be sensed are two mutually restrictive factors. Normally a bigger word length will lead to a lower sensing margin and make the TCAM structure less robustness to the noise effect. Therefore, it's crucial to increase the sensing margin as much as we can given the specific word length. Here we mathematically define the sensing margin as below:

        \begin{equation}
            \label{eq:sieq1}
            \beta = \frac{I_{m+1} - I_m}{I_m} = \frac{I_{m+1}}{I_m} - 1
        \end{equation}
        where $I_{m}$ is the largest possible current (or voltage after the sensing amplifier) of the closest match, and $I_{m+1}$ is the smallest possible current (or voltage after the sensing amplifier) of the next closest match. 
        A higher sensing margin $\beta$ indicates a fewer error in detecting the closest match, which is a function of a variety of parameters, including the memristor device ON/OFF conductance ratio $r=G_\text{ON}/G_\text{OFF}$, word length (the width of the array) $N$, the number of mismatches of the closest match $M$, the number of wildcard `X' in query vector/memory entry $K$, etc. 
        The following analyses aim to reveal the trade-off relationship among the parameters. 
        The effect of device variation and wire resistance is not taken into consideration for simplicity.
        % The 'match' case can be an exact match or the closest match with the smallest distance. Usually, a higher $\beta$ means that the 'match' case is much more distinguishable from counter parts. 
        % Before we perform the detailed analysis, we first introduce several characters that will be used. $V_s$: Search input voltage; $G_{on}$: Conductance of 'ON' state; $G_{off}$: Conductance of 'OFF' state; $r=\frac{G_{on}}{G_{off}}$: ON/OFF ratio of memristors; $N$: Word length, number of bits in total; $M$: Number of mismatch bits that we want to detect; $K$: Number of wild card 'X' in the query vector. 
        % We divide the situation into several cases:

        \begin{itemize}
            \item \textbf{Case 1}
        
            We start with the simplest case, where there is no wild card 'X' in the query vector, and we want to detect the exact match between the query and keys with a word length $N$. 
            The sense margin can be written as:
            \begin{align}
                \beta = \frac{I_{m+1}}{I_m} - 1 = \frac{G_{on}V_s + (N-1)G_{off}V_s}{NG_{off}V_s} - 1 = \frac{r-1}{N}
                \label{eq:sieq2}
            \end{align}
            where $V_s$ is the search voltage, and other parameters are defined previously. 

            One finds that the sensing margin increases with memristor ON/OFF ratio $r$ and decreases with the word length $N$. 
            On the other hand, if we want the sense margin to be at least $\beta_0$, the word length needs to be smaller than $\frac{r-1}{\beta_0}$. 
            For example, if the ON/OFF ratio ($r$) is 100, and the sense margin should be at least 0.5, the maximum word length $N_\text{max}$ is limited to $\frac{100 - 1}{0.5} = 198$.

            \item \textbf{Case 2}
            
            Instead of detecting the exact match, here in this case study, we detect the closest match (with $M$ mismatch bits).
            Wildcard `X' is also ignored in this case. 
            The sense margin can be written as:
            \begin{align}
                \beta = \frac{I_{m+1}}{I_m} - 1 & = \frac{(M+1)G_{on}V_s + (N-M-1)G_{off}V_s}{MG_{on}V_s + (N-M)G_{off}V_s} - 1 \\
                & = \frac{(M+1)r + N-M-1}{Mr+N-M} - 1 \\
                & = 1/\left(M + \frac{N}{r-1}\right)
            \label{eq:sieq3}
            \end{align}

            In addition to the trade-offs revealed in Case 1, one also finds that the sensing margin $\beta$ also decreases with the number of mismatch bits $M$.
            So, fewer number of mismatch bits is preferred for a given application. 

            % This case is more common when applying TCAM in the distance search application such as the memory attention in our MANN. This is because the query and keys cannot be exactly the same and there exits some randomness when generating the hashcodes. Therefore, even the closest distance can have a large number of mismatch bits $M$, resulting in a significant drop of sense margin. The word length $N$ and the on/off ratio $r$ can also affect the sense margin but the major factor is the number of mismatch bits $M$. One thing we can do to optimize the sense margin is to decrease the number of mismatch bits given a certain application. We will go through another analysis after.

            % \item \textbf{Case 3}

            % The query hashcode has $K$ wild card 'X' and we want to detect the exact match between the query and keys with word length of $N$. The sense margin can be written as:
            % \begin{align}
            %     \beta = \frac{I_{m+1}}{I_m} - 1 = \frac{G_{on}V_s+(N-K-1)G_{off}}{(N-K)G_{off}V_s} - 1 = \frac{r-1}{N-K}
            %     \label{eq:sieq4}
            % \end{align}
            % Since the search of 'X' won't contribute any current in the bit line (shown in Fig. \ref{fig:TCAM}\textbf{a}), the sense margin increase by a factor of $\frac{N}{N-K}$ comparing that with Case 1.
            % Moreover, the maximum word length $N$ given the sense margin $\beta$ is also enhanced by $K$ which becomes $\frac{r-1}{\beta}+K$.

            \item \textbf{Case 3}
            
            Here, we consider the case that query hash code has $K$ wildcard bit `X' from the TLSH step, and and we want to detect the closest distance with $M$ mismatch bits between the query and keys with word length of $N$. The sense margin can be written as: 
            \begin{align}
                \beta = \frac{I_{m+1}}{I_m} - 1 & = \frac{(M+1)G_{on}V_s + (N-M-K-1)G_{off}}{MG_{on}V_s + (N-M-K)G_{off}V_s} - 1 \\
                & = \frac{(M+1)r + N-M-K-1}{Mr+N-M-K} - 1 \\
                & = 1/\left(M + \frac{N-K}{r-1}\right)
            \label{eq:sieq5}
            \end{align}

            It is expected because introducing the wild card equivalently reduces the word length (ignoring the wire resistance effect). 
            From Equation \ref{eq:sieq5}, one concludes that the sense margin for the closest match increases with decreasing mismatch bits $M$ and wordlength $N$, and rising number of wildcard $K$ and device ON/OFF ratio $r$. 
            In addition, when we introduce `X' in hashcodes of the query, the number of mismatch bits $M$ of the nearest neighbor is usually reduced, further increasing the sensing margin. 
            % If we only see from Equation \ref{eq:sieq5}, the sense margin only increases a little bit since under most cases $r \gg K$. However, normally when we introduce 'X' in hashcodes of the query, the number of mismatch bits $M$ of the nearest neighbor is reduced. We use $P$ to represent the number mismatch bits reduced on account of 'X' in the query hashcode. Then the sense margin becomes: $\beta = \frac{1}{M-P+\frac{N-K}{r-1}}$. We can see that by decreasing $M$ we can always increase the sense margin. 
            % We need to clarify that all the techniques used to increase sense margin must be built on not degrading too much accuracy. 
        \end{itemize}

        From the above case study, we conclude that both decreasing the number of mismatch bits $M$ and increasing the number of the wild card bits $K$ improve the sense margin and robustness of the search operation. 
        The ternary locality sensitive hashing (TLSH) that we proposed in this work achieves both goals at the same time while maintaining the same accuracy. 
        The main reason behind this is that hashcodes generated by random projections have some redundancy bits that can be cast away by assigning them to the wildcard bit `X'. 
        Here we show the statistical results extracted from the few-shot classification to demonstrate the idea.

        \begin{figure}[!htbp]
            \centering 
            \includegraphics[width=0.95\textwidth]{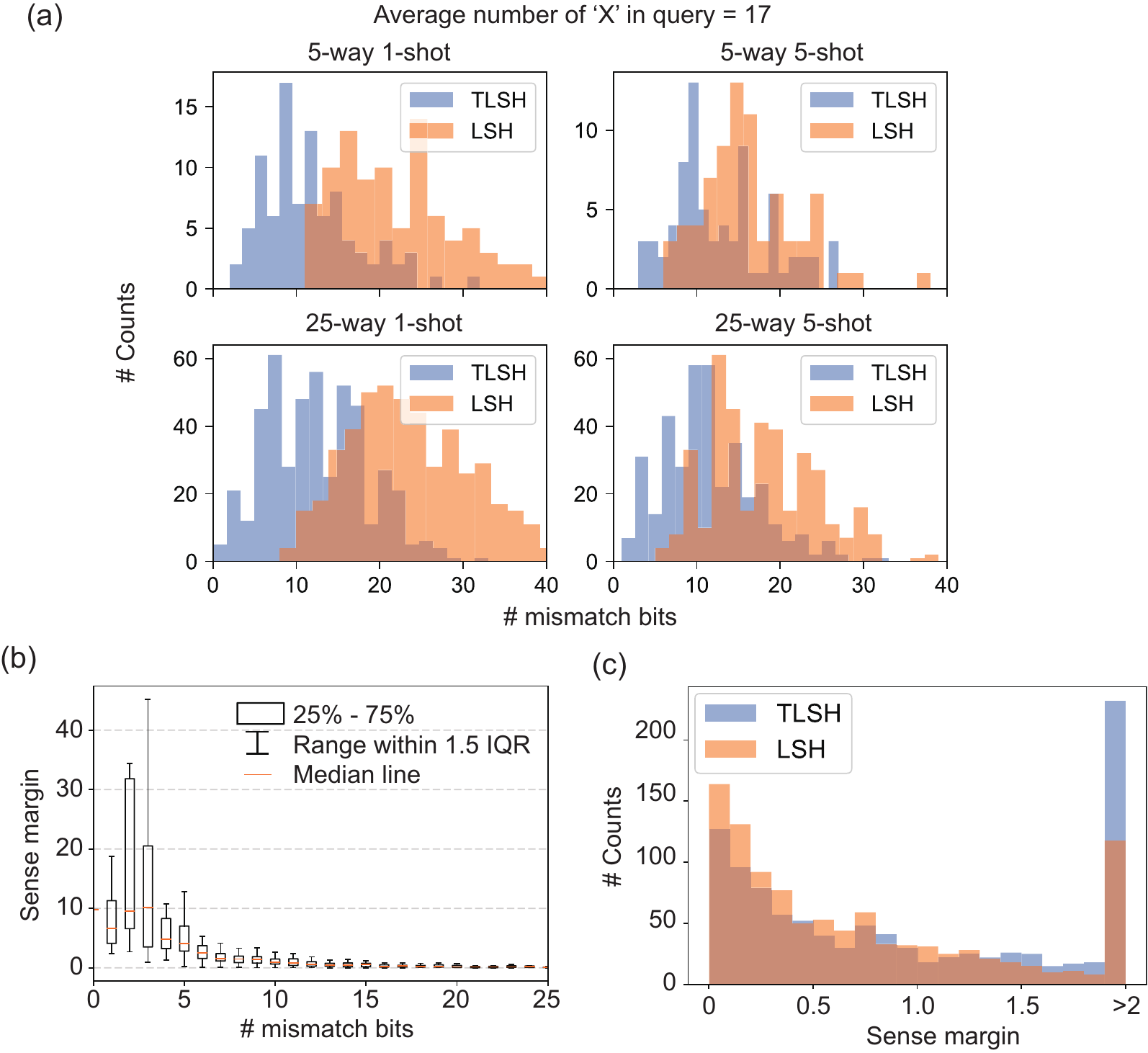}
            \caption{
            \rev{\textbf{Sense margin analysis of the few-shot classification on the Omniglot dataset.}
            \textbf{a,} Distribution of the mismatch bits of the closest match between query and keys in 4 different tasks. We can see that by using TLSH in the hashing operation, we can significantly decrease the number of mismatch bits.
            \textbf{b,} Relationship between the sense margin and the number of mismatch bits $M$ of the closest match measured in 4 tasks. The sense margin decreases as the $M$ increases.
            \textbf{c, } Distribution of the sense margin for TLSH and LSH. The probability that the sense margin is larger than 2 using TLSH method is twice as high as that using LSH. This verifies that our TLSH method can significantly increase the sense margin.
            }}
            \label{fig:sense_margin}
        \end{figure}

        We first analyze the statistics of the mismatch bits numbers (or hamming distance) of the closest match $M$ during the few-shot learning experiments. 
        The distribution of the hamming distance is shown in Supplementary Fig. \ref{fig:sense_margin}\textbf{a}. 
        The result shows that by introducing 'X' in the query hashcodes, the number of mismatch bits $M$ of the closest match decreases significantly. 
        The experimental sensing margin is then extracted from the output current of the closest match (the smallest current $I_0$) and the next close match (the 2nd smallest current $I_1$), by the equation:
        $\beta = \frac{I_1}{I_0} - 1$.
        As expected, the experimental sensing margin is inversely proportial to the the hamming distance, \ie $\beta \propto \frac{1}{M}$ as is shown in Supplementary Fig. \ref{fig:sense_margin}\textbf{b}.
        Though the sense margin here is not the worst sense margin characterized by Equation \ref{eq:sieq5}, we still see a similar trend. 
        As a result, the sensing margin is improved by introducing the wildcard `X' from the TLSH because of the reduced hamming distance of the closest match. 
        The conclusion is clearly demonstrated in the experimental sensing margin distribution shown in Supplementary Fig. \ref{fig:sense_margin}\textbf{c}. 
        Specifically, the probability that the sense margin is larger than two with TLSH is more than two times higher than that using LSH. 
        In light of the above analysis, we can conclude that the reduced mismatch bit number from our TLSH scheme improves the sense margin of the crossbar-based TCAM for the closest match search.

        \begin{figure}[!htbp]
            \centering 
            \includegraphics[width=0.95\textwidth]{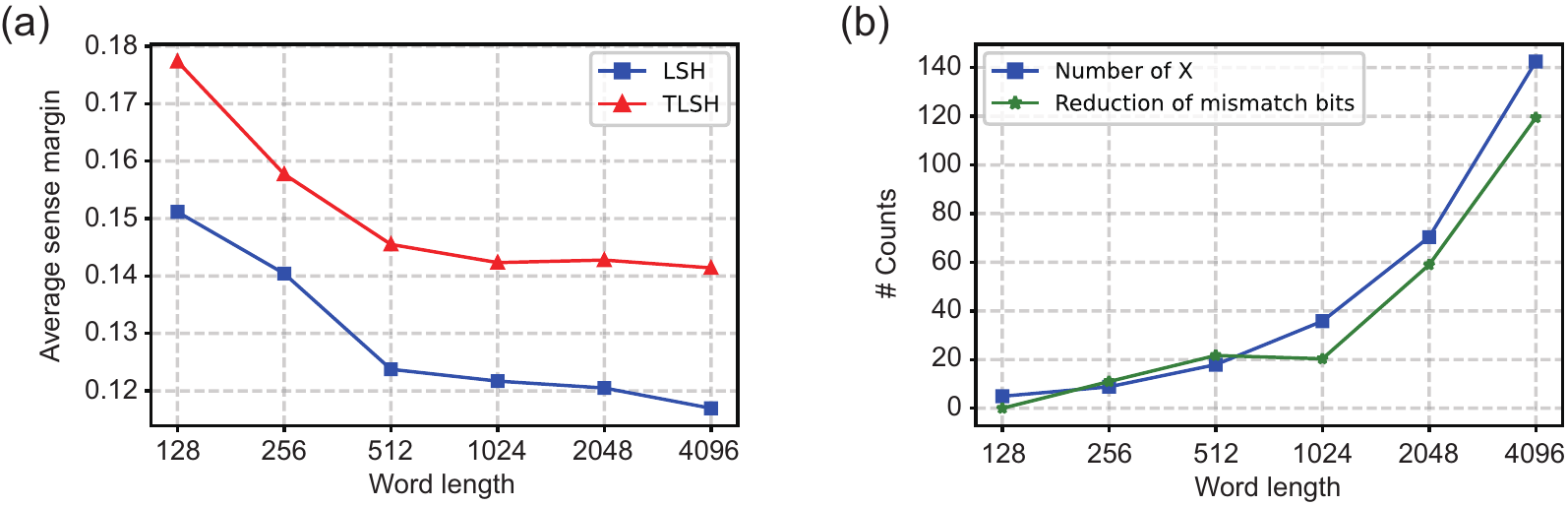}
            \caption{
            \rev{\textbf{Sense margin analysis of the 5-way 1-shot classification on the Mini-ImageNet dataset.}
            \textbf{a, } Average sense margin drops as the word length increases. Sense margin on the hashcodes generated by TLSH method is always higher than that by LSH method. 
            \textbf{b, } Reduction of mismatch bits grows with the word length and follows the similar trend as the number of 'X' in the query.
            }
            }
            \label{fig:sense_margin_imagenet}
        \end{figure}

        Similar phenomena can also be observed in a scaled problem for few-shot learning with Mini-ImageNet dataset. 
        The result in Supplementary Fig. \ref{fig:sense_margin_imagenet}\textbf{a} shows that the sensing margin decreases with the word length $N$, and the sensing margin with TLSH is always higher than that with LSH, because of the introduced wildcard bit `X'.
        Supplementary Fig. \ref{fig:sense_margin_imagenet}\textbf{b} also show that the introduction of the wildcard `X' also reduces the number of mismatch bits, further improving the sensing margin. 
        % To evaluate the scalability of our TLSH plus TCAM method, we simulate the sense margin on the Mini-ImageNet dataset as shown in Supplementary Figure. \ref{fig:sense_margin_imagenet}. 
        % We can see that with the same accuracy, TLSH can reduce the mismatch bits of the nearest neighbor and thus increase the sense margin. \can{Fig. S16a suggest otherwise? The hamming distance is higher for TLSH???}
        % We can also find a similar trend for the growth of reduction of mismatch bits and the number of wild card 'X'.
        % From the above analysis we can conclude that our TLSH plus TCAM method can provide a better sense margin when executing distance search tasks compared with conventional LSH techniques\citesicite{Ni2019NE}. 
        
        Note that our TLSH method can also combine with other TCAM structures like 2T-2R\citesicite{li20131_si, li2021sapiens_si}, 2Flash\citesicite{fedorov2014area_si} 2FeFET\citesicite{Ni2019NE_si, Laguna2019GLSVLSI_si} which have the similar characteristic, but use latched voltage signal as the output. 
        As mentioned, the sense margin of our crossbar-based TCAM can be further improved with the 3-bits encoding method, as discussed in Supplementary Fig. \ref{fig:3dpe}. 
        % In the 3-bits encoding method, the `match' case contributes a smaller current to the match line than 2-bit encoding, at a  sacrifices the bit density. 
        % Therefore, it's better for those devices with a relatively lower on/off ratio like MRAM\citesicite{apalkov2016magnetoresistive}.

    \label{sinote:sense_margin}
    \end{sinote}
    }

    \begin{sinote} 
        \textbf{Energy and Latency estimation}
    % \section*{Energy and Latency estimation}

    The latency and energy consumption estimation for end-to-end MANN on the Omniglot dataset is based on the experimental data extracted from our integrated crossbar system. For the latency estimation in CNN layers, we considered channel-wise parallelism, where every channel in a convolutional layer is working in a parallel way. Therefore, the latency of CNN layers for every input image is depended on the number of input vectors and read time through column wires. Each read time for vector-matrix-multiplication in the latency estimation is \SI{10}{\nano\second}. From the CNN structure we can get the total time latency for CNN layers:
    $$
    T = t_{read}\times N_{input} = (784+784+196+196)\times10 = \SI{19.6}{\micro\second}
    $$
    For the latency in TLSH and TCAM, since we need to predefine the memory size for the few-shot learning task, \ie, 256 entries for Omniglot, we still need to partition the hashing matrix and memory matrix into multiple $64\times64$ arrays, which is the size in the experiment. In this case, we still need to consider the time latency of full adders for aggregating the results from partitioned arrays. We use the \SI{2.5}{\nano\second} latency of conventional 16T 8-bit adder reported in ref\citesicite{ghadiry2013dlpa_si}. Therefore, we get the time latency for TLSH and TCAM: $\SI{12.5}{\nano\second}+\SI{12.5}{\nano\second}=\SI{25}{\nano\second}$ 

    The energy consumption estimation for Omniglot is done by reading the conductance matrix out throughout the experiment. The total energy is calculated by:
    \rev{
    $$
    E=\sum_{i,j}V_{ji}^2 G_i t_{\text{pulse}}
    % E=\sum_{i,j}V_{ji}^2 \odot G_i 
    $$
    where $V_{ji}$ is the $j$th input voltage vector to $i$th conductance matrix, $G_i$ is the $i$th readout conductance matrix and $t_{\text{pulse}}$ is the pulse width of the input}. The final energy consumption of the entire MANN per image for the 5-way 1-shot problem is \SI{1.26}{\micro\joule} where search energy only consumes \SI{13}{\pico\joule} (including TLSH). We acknowledge that the energy and latency overhead of peripheral circuits which need to realize the entire MANN is not taken into account yet since it's not fully optimized and it's difficult to estimate. The numbers here mainly give an estimation for the implementation in crossbar arrays. 

    For the estimation for the Mini-ImageNet dataset, we mainly focus on the overhead in the search operation which is the key point in this work. For latency estimation, according to our architecture for scaled-up networks, time won't increase since the TLSH and TCAM are all working in a parallel way which is \SI{25}{\nano\second} based on our aforementioned assumption. 
    For the energy consumption for Mini-ImageNet on the 5-way 1-shot task, we follow the methods above while we use the simulated conductance matrix according to our measurement. The final energy consumption for a single image search across the entire memory module is \SI{594}{\pico\joule}.

    In summary, we estimated about \SI{1.26}{\micro\joule} per Omniglot image inference (5-way 1-shot) with our full crossbar-based MANN \rev{with $>$1,000$\times$ improvement over GPU (\SI{1.38}{\milli\joule} on GPU).}  
    Different from MANN implemented on a conventional GPU backed by DRAM, which spends most energy and time on memory transfer in the memory search, both the energy consumption and time latency in crossbar arrays for MANN mainly comes from the CNN controller (\SI{19.6}{\micro\second} for time latency and \SI{12.5}{\micro\joule} for energy consumption). 
    If we consider the search operation only, the consumed energy for one search operation is only \SI{2.1}{\pico\joule} for the Omniglot dataset (128 bits), \rev{$~$2857$\times$ improvement over GPU.}
    For Mini-ImageNet dataset, we achieve \SI{82.4}{\pico\joule} per search (4,096 bits), over $5.1\times10^4$ improvement to GPU on the same task\citesicite{kazemi2021flash_si}.
    \rev{
    The GPU latency numbers reported here are acquired using PyTorch Profiler\citesicite{paszke2019pytorch_si}, and the energy consumption numbers from the NVIDIA System Management Interface (nvidia-smi).}
    
    Those numbers can be further improved by choosing smaller input voltages and optimizing peripheral circuits with more advanced technology nodes. 
    \rev{
    Here, we demonstrate that energy consumption can be improved by simply voltage scaling. 
    In the experiment, we reduce the search voltage by a factor of 10: from \SI{0.2}{\volt} to \SI{0.02}{\volt}. 
    In this case, the power consumption is lowered 100$\times$ with minor accuracy loss (from 76.8\% to 77.5\% for the 25-way 1-shot problem). 
    It is noteworthy that the efficiency is not necessarily improving with larger crossbars\citesicite{wan2021edge_si}. We found the size of 64$\times$64 or 128$\times$128 is a sweet spot when considering the output current, matrix utilization rate, etc. Quantitively, we compared search energy in the distance search
    operation with different approaches and listed them in the table below, clearly showing our approach's benefit and scalability. The result of the search energy is averaged over 11,875 search operations in the 25-way 1-shot task based on the experimentally measured data. 
    To make a fair comparison, we report the search energy per bit of our approach and achieve the lowest energy consumption over previous work.
    }

    \begin{table}[H]
        \centering
        \caption{\rev{Energy consumption per bit per search operation of TCAM}}
        \begin{tabular}{l|lll}
        \hline
         & Ref.\citesicite{karunaratne2021robust_si} & Ref.\citesicite{Ni2019NE_si} & Our work\\
        \hline
        Energy consumption per bit per search/fJ & 2.5 & 0.4 & 0.17 \\
        \hline
        \end{tabular}
        \label{tb:energy_compare}

    \end{table}

    \label{sinote:energy}
    \end{sinote}
    
    % \section{Supplementary Discussions}
    % \section{Supplementary Methods}
    % \section{Supplementary References}
    % \label{not:si_searching}
    
    \section{Supplementary References}
    \bibliographystylesicite{naturemag}
    \bibliographysicite{sicite}
    
    \end{document}